\documentclass[%
 reprint,
superscriptaddress,
 amsmath,amssymb,
 aps,
prb,
]{revtex4-2}

\usepackage{graphicx}% Include figure files
\usepackage{dcolumn}% Align table columns on decimal point
\usepackage{bm}% bold math

\usepackage{bbold}

\usepackage[normalem]{ulem}
\usepackage{cancel}

\usepackage{natbib}

\usepackage[dvipsnames]{xcolor}

\newcommand{\red}{\textcolor{Black}}      % For the modified text
\newcommand{\green}{\textcolor{Black}}      % For the modified text 
\begin{document}

\preprint{APS/123-QED}

\title{Spin-orbit interaction in tubular prismatic nanowires}

\author{Anna Sitek}
\affiliation{Institute of Theoretical Physics,
	Wroclaw University of Science and Technology,
	Wybrze{\.z}e Wyspia{\'n}skiego 27, 50-370 Wroclaw, Poland.}

\author{Sigurdur I.\ Erlingsson}%
%\author{Vidar Gudmundsson}
%\affiliation{Science Institute, University of Iceland, Dunhaga 3, IS-107 Reykjavik, Iceland}

\affiliation{Department of Engineering, Reykjavik University, Menntavegur 1, IS-102 Reykjavik, Iceland.}

\author{Andrei Manolescu}
\affiliation{Department of Engineering, Reykjavik University, Menntavegur 1, IS-102 Reykjavik, Iceland.}

%\date{\today}

\begin{abstract}
We theoretically study the spin-orbit interaction in the outer regions of core-shell nanowires that can act as tubular, prismatic conductors. The polygonal cross section of these wires induces non-uniform electron localization along the wire \red{perimeter}. In particular, low-energy electrons accumulate in the corner regions, and in the case of narrow shells, conductive channels form along the sharp edges. In contrast, higher-energy electrons are shifted toward the facets. These two groups of states may be separated by large energy gaps, which can exceed the room-temperature energy in the case of triangular geometries. We compare the impact of spin-orbit interaction on the corner and side states of hexagonal and triangular shells grown on hexagonal cores as well as on triangular shells grown on triangular cores. We find that the spin-orbit
splitting, and thus the degeneracy of energy states at finite wave vectors, strongly depend on the tube's geometry. We demonstrate that the weak spin-orbit coupling observed in clean wires can be significantly enhanced if the intermixing of core and shell materials takes place. 
\red{
Moreover, we show that the energy spectrum in the presence of spin-orbit interaction allows for estimating the interaction between states and shows that triangular shells can act as three independent wires in the low-energy regime, while they behave as interacting systems at higher-energy ranges.
}
\end{abstract}

\maketitle

\section{\label{sec:intro}Introduction}

Spin-orbit interaction (SOI) refers to the coupling between an electron's spin and the magnetic field induced by the electron's motion in an electrostatic field. In solid-state systems, the electrostatic field, and thus the SOI, arises from bulk (Dresselhaus SOI) and structural (Rashba SOI) inversion asymmetries. The Dresselhaus SOI is an intrinsic material property resulting from the crystallographic structure, whereas the Rashba SOI is governed by the confining potential and can be tuned externally through electrostatic fields or controlled during the growth process. SOI plays a crucial role in the operation of spin-orbit qubits \cite{Nadj_Perge10, van_den_Berg13} and devices such as spin transistors \cite{Koo09, Wojcik14} and spin filters \cite{Ngo10, Kohda13, Wojcik17}. Additionally, in the presence of a magnetic field and a superconducting materials, SOI facilitates the formation of Majorana states at the ends of semiconducting wires \cite{Lutchyn10, Oreg10, Stanescu17, Laubscher21}. Most proposals focus on single-material systems, such as InSb or InAs, to host Majorana states, where the internal geometry of the wires is typically not considered. In contrast, our group has proposed the use of the outer regions of the core-shell nanowires to host Majorana states \cite{Manolescu17, Stanescu18}.

Core-shell nanowires are radial heterojunctions formed by overgrowing a quantum wire with layers of different materials. Although cylindrical structures have been obtained \cite{Kim17}, in most cases the crystallographic properties of the components lead to polygonal cross sections. The most common are hexagonal nanowires \cite{Blomers13, Rieger12, Haas13, Fickenscher13, Funk13, Jadczak14, Gul14, Weiss14b, Erhard15, Shi15, Sonner19}, but triangular \cite{Qian04, Qian05, Baird09, Heurlin15, Dong09}, rectangular \cite{Guniat19, Fonseka19}, and even dodecagonal \cite{Rieger15} shapes have also been demonstrated. Furthermore, nanowires combining two different polygons within a single cross section are technologically feasible. Specifically, triangular shells can be grown on hexagonal cores such that either the vertices of the internal and external boundaries match \cite{Yuan15}, or their sides are parallel \cite{Goransson19}.

The combination of \red{two semiconductors with significantly different band gaps} can result in a type I band alignment at the junction between the materials, leading to the transfer of electrons to the material \red{with the narrower gap}. By properly adjusting the material and geometric parameters, it becomes possible to create a radial heterojunction in which electrons are confined in the core region \cite{Bertoni11,Royo13,Jadczak14} or in the shell region \cite{Blomers13,Fickenscher13,Sonner19,Funk13}.
In terms of electron localization, the former case resembles single-material wires, where the electron ground state is localized at the center of the sample \cite{Bertoni11,Wong11}. As a result, the shape of the cross section does not significantly affect the properties of the electron in its lowest-energy state. For higher-energy states, the probability distributions exhibit several maxima, which are determined by the number of vertices. However, these maxima often overlap, causing electrons excited to higher energies to remain near the center of the cross section.
As a result, the properties of electrons confined in the cores of core-shell wires or in single-material wires, and in particular the SOI, can be effectively described using a cylindrical wire model \cite{Woods19}. However, this description does not apply when the electrons occupy the shell region. The electron distribution in the shell differs significantly between the circular and polygonal cross sections. In cylindrical wires, the electrons are uniformly distributed along the circumference of the cross section. In contrast, for prismatic tubes, the electrons tend to localize in different regions of the cross section, depending on their energy.

In the present paper, we consider core-shell nanowires with hexagonal and triangular cross sections and calculate the effects of \red{SOI} 
\red{arising from the extended band-offset potential and an external electric field}
on electronic energy spectra using the $\bm{k}\!\cdot\!\bm{p}$ method. 
\red{
We show that the SOI-induced degeneracy differs between the hexagonal shell and the other geometries studied. Moreover, the effect of SOI is significantly enhanced in structures where intermixing of the core and shell materials occurs. In the presence of SOI, the corner states of triangular shells resemble the spectrum of three independent wires, while the lowest side-localized states exhibit clear signatures of interaction.
}

\red{The paper is organized as follows.}
Section II introduces the model, while Sec. III presents the results for different geometries. Conclusions are summarized in Sec. IV.

%%%---------------------------------------------------------------- 

\section{\label{sec:model}Model}

\begin{figure}
    \centering
    \includegraphics[scale=0.67]{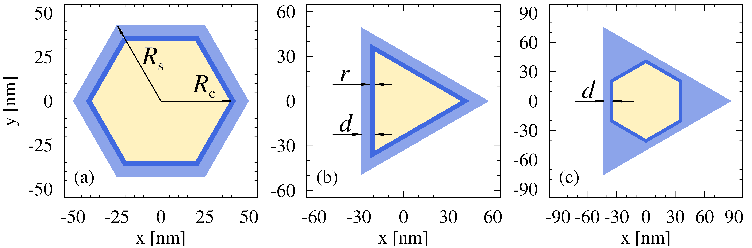}
    \caption{Cross sections of (a) hexagonal, (b) triangular, and (c) Tri-Hex wires. $R_{\mathrm{c}}$ and $R_{\mathrm{s}}$ represent the external radii of the core and shell (wire), respectively. The shell thickness is denoted by $d$ and the intermixing length is indicated by $r$. $R_{\mathrm{c}}=40$ nm, $d=8$ nm, and $r=2$ nm.}
    \label{fig:SOI_Samples}
\end{figure}

We study the electrons confined in the shells of infinitely long InP-InAs core-shell nanowires with three different cross-sectional geometries: a triangular shell grown on a triangular or hexagonal core (Tri-Hex), and a hexagonal shell covering a hexagonal core, as shown in Fig.\ \ref{fig:SOI_Samples}. 
\red{
Our computational approach can be applied to core-shell nanowires with shells that have the zinc-blende crystallographic structure and can be easily modified to describe structures with wurtzite shells. We chose InP and InAs materials because both Tri-Hex \cite{Goransson19} and hexagonal \cite{Mohan06} InP-InAs core-shell nanowires have been demonstrated. Moreover, triangular InP nanowires \cite{Kriegner11}, which could be used as cores, have also been experimentally realized.
}
%

%%%%%%%%%%%%%%%%%%%%%%%%%%%%%%%%%%%%%%%%%%%%%%%%%%%%%%%%%
%%%%%%%%%%%%%%%%%%%%%%%%%%%%%%%%%%%%%%%%%%%%%%%%%%%%%%%%%

We consider wires grown along the crystallographic direction [111], for which the Dresselhaus SOI can be neglected \cite{Luo11}.
The eight-band Kane Hamiltonian is given by 
\begin{eqnarray}
	H = \left[
	\begin{array}{cc}
		H_{cc} & H_{cv} \\
		H_{cv}^{\dagger} & H_{vv} \\		
	\end{array}\right]		\, ,
\end{eqnarray}	
where $H_{cc}$ and $H_{vv}$ are diagonal matrices $2\times 2$ and $6\times 6$, respectively, describing the two lowest conduction bands and the six highest valence bands. Assuming the wires are grown along the $z$ axis, the conduction-band Hamiltonian is
\begin{eqnarray}
	H_{cc} = \left(\frac{\bm{p}^2}{2m_0} + V\!(x,y) \right)	\mathbb{1}_{2\times 2} \, ,
\end{eqnarray}	
where
\begin{eqnarray}
\bm{p} = \left(-i\hbar\frac{\partial}{\partial x}, -i\hbar\frac{\partial}{\partial y}, \hbar k_z \right) 
\end{eqnarray}
is the momentum operator, $k_z$ represents the wave vector in the growth direction, and $m_0$ stands for the free-electron mass.   
The potential energy
\begin{eqnarray}	
	V\!(x,y) = V_{\mathrm{BO}}(x,y) + V_{\mathrm{E}}(x,y)
\end{eqnarray}
includes the contributions due to conduction-band offset which penetrates into the shell [$V_{\mathrm{BO}}(x,y)$] and the potential due to the external electric field [$V_{\mathrm{E}}(x,y)$].

We assume that initially the intermixing of the core and shell materials, InP and InAs, takes place. However, the proportion of core material gradually decreases while that of the shell material increases during the shell growth process \cite{Fabian24}.
\red{In our model, the volume in which the contributions from both materials are present is defined by its cross section, i.e., a ring surrounding the core (darker blue regions in Fig.\ \ref{fig:SOI_Samples}). The thickness of this ring is denoted by $r$  [Fig.\ \ref{fig:SOI_Samples}(b)] and referred to as the intermixing length. In general, $r$ can range from zero to the arbitrary value, but in }
this study, we assume a minimal intermixing length of $r=2$ nm. This controlled material mixing results in a tilted potential within the shell, which corresponds to an extended band-offset potential ($V_{\mathrm{BO}}$) that (in our model) decreases linearly 
\red{ along the radial lines of the grid}
from a value of $V_{\mathrm{CBO}} = 656.6$ meV \cite{Vurgaftman01} at the core-shell boundary, which is the conduction-band offset at the InP-InAs heterojunction.
%%
%\begin{eqnarray}	
%	V_{\mathrm{BO}}(x,y) = V_{\mathrm{CBO}} - \frac{V_{\mathrm{CBO}}}{r} \sqrt{x^2 + y^2}  \, .
%\end{eqnarray}
%

If the distance $r$ is less than the thickness of the shell (\red{$d$ in} Fig. \ref{fig:SOI_Samples}), the potential \red{$V_{\mathrm{BO}}$} drops to zero, i.e., reaches the bottom of the shell material conduction band at \red{the external boundary of the polygon defined by $r$, i.e., within the shell, and the outer region of the shell consists solely of InAs.} If $r\geqslant d$ the entire shell contains \red{P atoms} and the value of potential $V_{\mathrm{BO}}$ at the wire external boundary increases with $r$, but stays within the interval $(0, V_{\mathrm{CBO}})$.
This potential gradient creates a built-in, or intrinsic, radial electric field in the wire, thereby inducing SOI. Furthermore, we consider the possibility of an external electric field ($V_{\mathrm{E}}$) applied perpendicular to the wire, which can alter the electron distribution within the shell \cite{Sitek15} and further influence the SOI.

Taking the zero-energy level at the bottom of the conduction band of the shell \red{and neglecting the off-diagonal matrix elements}, the valence-band Hamiltonian is given by 
\begin{eqnarray}
	H_{vv} = \mathrm{diag}\left[
\begin{array}{c}	
	V(x,y) - E_g, \\
	V(x,y) - E_g , \\
	V(x,y) - E_g , \\
	V(x,y) - E_g,  \\
	V(x,y) - E_g - \Delta_{\mathrm{so}},  \\
	V(x,y) - E_g - \Delta_{\mathrm{so}}   
\end{array}	 
	 \right] \, ,
\end{eqnarray}	
where $E_g$ is the semiconductor band gap and $\Delta_{\mathrm{so}}$ is the spin-split gap.

The off-diagonal block is  
\begin{eqnarray}
	&& H_{cv} = \frac{P_0}{\hbar} \times \\ && \left[
	\begin{array}{cccccc}	
		-\frac{1}{\sqrt{2}}p_+ & \sqrt{\frac{2}{3}}p_z & \frac{1}{\sqrt{6}}p_- & 0 & -\frac{1}{\sqrt{3}}p_z & -\frac{1}{\sqrt{3}}p_-   \\		
		0 & -\frac{1}{\sqrt{6}}p_+ & \sqrt{\frac{2}{3}}p_z & \frac{1}{\sqrt{2}}p_- & -\frac{1}{\sqrt{3}}p_+ & \frac{1}{\sqrt{3}}p_z	 
	\end{array}
	\right] \nonumber \ ,
\end{eqnarray}
where 
\begin{eqnarray}
	P_0 &=& \frac{\hbar}{m_0} \langle S\sigma|p_x|X\sigma \rangle 
	    = \frac{\hbar}{m_0} \langle S\sigma|p_y|Y\sigma \rangle  \\
	    &=& \frac{\hbar}{m_0} \langle S\sigma|p_z|Z\sigma \rangle	 \nonumber
\end{eqnarray}
is proportional to the nonvanishing momentum matrix elements between the conduction ($|S\sigma \rangle$) and the valence $(|X(Y,Z)\sigma \rangle$) band-edge Bloch states at the $\Gamma$ point of the Brillouin zone. Here, $\sigma$ represents spin and $p_{\pm} = p_x \pm ip_y $.

The folding-down transformation \cite{Fabian07, Winkler} allows us to obtain the reduced Hamiltonian that describes only the conduction electrons, but with the influence of the highest valence bands on their energies. For the shell, this Hamiltonian takes the form
\begin{eqnarray}
\label{HSOI}
\mathcal{H}_{cc} &=& H_{cc} - H_{cv}\left(H_{vv}-E\right)^{-1}H_{vc}  = \\
&=&   
\left\{\frac{\hbar^2}{2m^{*}}\left(  k_{z}^2  
-\frac{\partial^2}{\partial x^2}  
-\frac{\partial^2}{\partial y^2} \right)  
 + V(x,y)
\right\} \mathbb{1}_{2\times 2}  \nonumber \\ 
&&\hspace{0.5 cm} + k_z \bigg(\alpha_x(x,y)\sigma_x + \alpha_y(x,y)\sigma_y\bigg)  \nonumber \, ,
\end{eqnarray}
where
\begin{eqnarray}
\frac{1}{m^*} \approx \frac{1}{m_0}+\frac{2P_0^2}{3\hbar^2}\left(\frac{2}{E_g}+\frac{1}{E_g+\Delta_{so}} \right)
\label{m_eff_z}
\end{eqnarray}
is the effective mass, and the Rashba spin-orbit coefficients are
\begin{subequations}
\label{alpha_xy}
\begin{equation}	
   \alpha_x(x,y) = -\alpha_0 \frac{\partial}{\partial y}V(x,y) \, ,  \\
\end{equation}
\begin{equation}
	 \alpha_y(x,y) =  \alpha_0 \frac{\partial}{\partial x}V(x,y) \, ,
\end{equation}
\end{subequations}
where
\begin{eqnarray}
\label{alpha0}
	 \alpha_0 = \frac{P_0^2}{3} \left(\frac{1}{E_g^2} - \frac{1}{\left(E_g+ \Delta_{so}\right)^2} \right) 
\end{eqnarray}
is the amplitude of the Rashba coefficients, and $\sigma_x$ and $\sigma_y$ are the Pauli matrices \cite{Wojcik18,Wojcik19,Wojcik21}.  
The expressions for the effective mass [Eq. (\ref{m_eff_z})] and the Rashba coefficients' amplitude [Eq. (\ref{alpha0})] were derived based on the assumption that the semiconductor band gap of the shell ($E_g=417$ meV) is the largest energy within the studied subsystem. 
Note that for the narrow shells, the values of $m^*$ and $\alpha_0$ [Eqs.\ (\ref{m_eff_z}) and (\ref{alpha0}), respectively] might get slightly modified\textemdash  but this effect is beyond our present approximation since our goal is to establish the qualitative picture of the SOI in core-shell nanowires, in particular to specify the degeneracy of the states.

To obtain the energy dispersion, we use a discretization method developed in Ref.\ \onlinecite{Sitek18}. We construct a polar grid, on which we superimpose polygonal constraints \red{corresponding to the cross-sectional} geometry of the shell, and diagonalize the Hamiltonian [Eq.\ (\ref{HSOI})] for a large set of $k_z$ values, one at a time, on the reduced grid located within the shell. We apply the Dirichlet \red{and the infinite potential well boundary} conditions  at the shell-vacuum and core-shell polygonal boundaries \cite{Sitek15, Sitek16, Sonner19}. \green{To compute the transverse states, we numerically diagonalize the Hamiltonian in Eq.\ (\ref{HSOI}) \red{for \(k_z=0\)} using the effective mass of the conduction band [Eq. (\ref{m_eff_z})] \cite{Sitek15, Sitek16, Sonner19}. } This method allows us to model wires with arbitrary cross-sectional shapes, including nonsymmetric cases \cite{Sitek15, Sitek19}.

Our approach includes the geometrical confinement of the electrons inside the narrow tubular shells, but we neglect the effects of the Coulomb repulsion between electrons, typically represented by the Hartree potential, to be obtained with the Poisson equation.  Because we consider a low concentration of electrons confined in a narrow shell of 8\textendash24 nm thickness, the redistribution of electrons due to electrostatic forces is minimal, but computationally demanding. Nevertheless, we tested a couple of situations and indeed the corrections to the energy spectra due to the Hartree potential were not significant.

%%%---------------------------------------------------------------- 

\section{\label{sec:results}Results}

\red{
The numerical simulations presented below were performed for cross sections consisting of over 6000 grid points for hexagonal shells, over 7000 for triangular wires, and between 6200 and 7200 for Tri-Hex structures. To preserve cross-sectional symmetry, the number of angular grid points was chosen to be a multiple of the number of internal vertices. The material parameters were either taken directly from or calculated based on Ref.~\onlinecite{Vurgaftman01}. For the InAs shell, these parameters are: $E_g = 417$ meV, $\Delta_{\mathrm{so}} = 390$ meV, and $P_0 = 9\times 10^{-10}$ eVs. For InP, the band gap is $E_g = 1423.6$ meV. The InP-InAs valence-band offset is $350$ meV, and the resulting conduction-band offset is $V_{\mathrm{CBO}} = 656.6$ meV.
}

%%%%%%%%%%%%%%%%%%%%%%%%%%%%%%%%%%%%%%%%%%%%%%%%%%%%%%%%%
\begin{figure}
    \centering
    \includegraphics[scale=0.67]{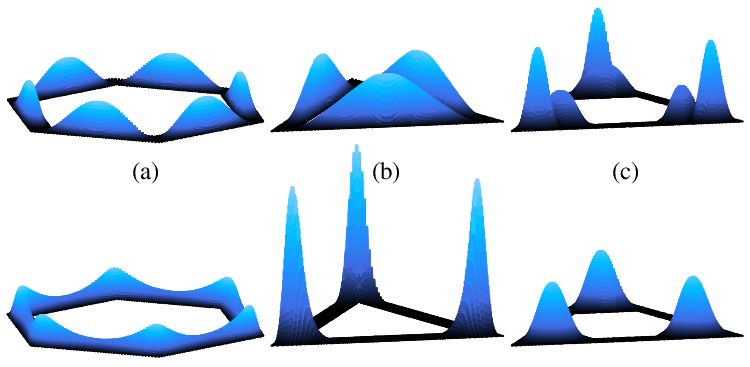}    
    \caption{Probability distributions corresponding to the lowest energy level (lower row) and the first energy level above the gap $\Delta$ (upper row) for (a) hexagonal, (b) triangular, (c) and Tri-Hex cross sections ($d=8$ nm, $R_{\mathrm{c}}=40$ nm).}
    \label{fig:L}
%\end{figure}
%%%%%%%%%%%%%%%%%%%%%%%%%%%%%%%%%%%%%%%%%%%%%%%%%%%%%%%%%
%%%%%%%%%%%%%%%%%%%%%%%%%%%%%%%%%%%%%%%%%%%%%%%%%%%%%%%%%
%\begin{figure}
	\centering
	\includegraphics[scale=0.67]{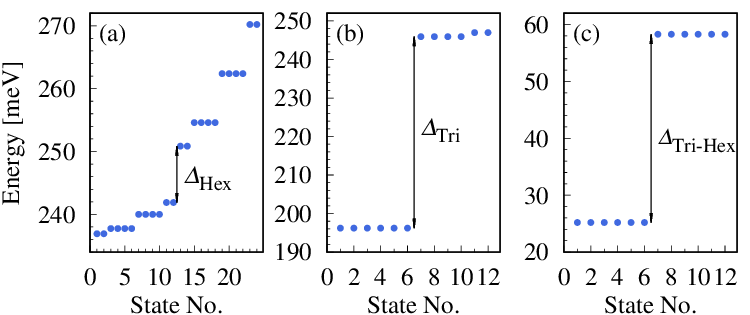}    
	\caption{Low-energy transverse states for the cross sections shown in Fig.\ \ref{fig:SOI_Samples}. The gap $\Delta$ separating the lowest (corner) states is: $\Delta_{\mathrm{Hex}}= 6$ meV, $\Delta_{\mathrm{Tri}}= 33$ meV, and $\Delta_{\mathrm{Tri\textrm{-}Hex}}= 31$ meV.}
	\label{fig:energy}    
\end{figure}
%%%%%%%%%%%%%%%%%%%%%%%%%%%%%%%%%%%%%%%%%%%%%%%%%%%%%%%%%

%%%%%%%%%%%%%%%%%%%%%%%%%%%%%%%%%%%%%%%%%%%%%%%%%%%%%%%%%
%--- Energy kz without SOI ------------------------------ 
\begin{figure*}
	\centering
	\includegraphics[scale=0.67]{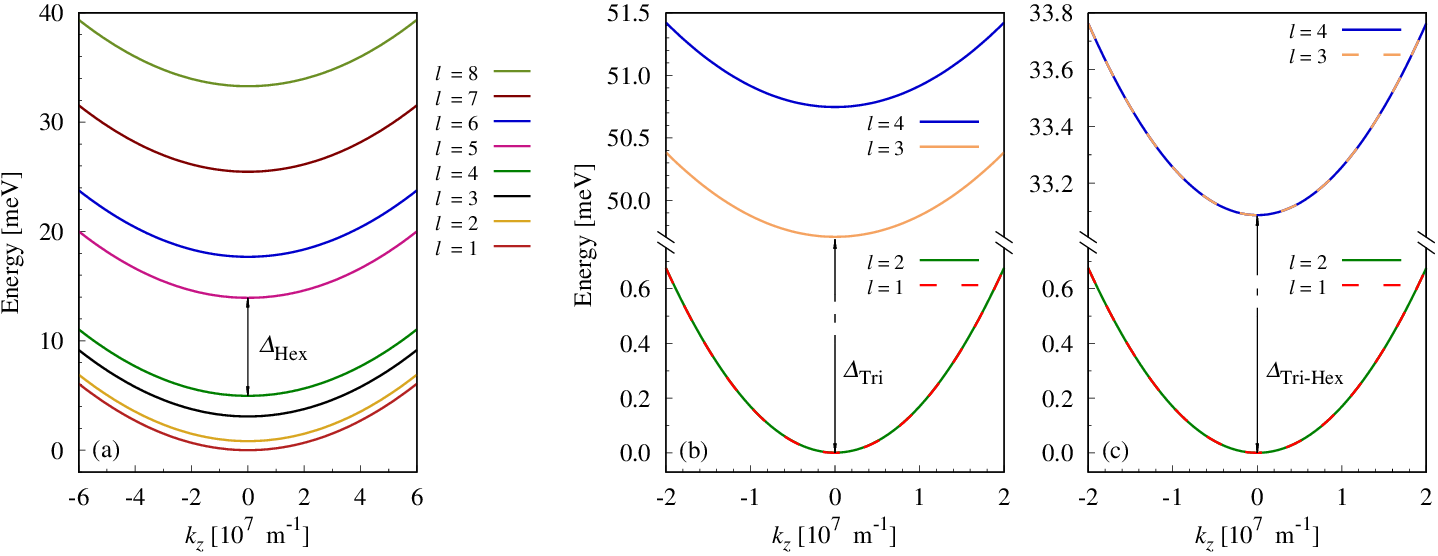}    
	\caption{Low-energy states for the cross sections shown in Fig.\ \ref{fig:SOI_Samples} without SOI. \red{The energies are shifted with respect to the ground state energy at $k_z=0$. (a) States $l=1$, $l=4$, $l=5$, and $l=8$ are twofold degenerate, while states $l=2$, $l=3$, $l=6$, and $l=7$ are fourfold degenerate. (b) and (c) States $l=1$ and $l=4$ are twofold degenerate, while states $l=2$ and $l=3$ are fourfold degenerate.} }
	\label{fig:energy_kz} 
\end{figure*}
%%%%%%%%%%%%%%%%%%%%%%%%%%%%%%%%%%%%%%%%%%%%%%%%%%%%%%%%%

\subsection{Transverse states}

Below, we summarize the properties of the hexagonal, triangular, and Tri-Hex quantum rings, which represent the cross-sectional areas of the conductive shells (Fig.\ \ref{fig:SOI_Samples}). In the triangular and hexagonal systems, the cross sections of the core alone and the entire core-shell wire are the same polygons [Figs.\ \ref{fig:SOI_Samples}(a) and \ref{fig:SOI_Samples}(b)]. The Tri-Hex system combines two distinct geometries: a hexagonal core with a triangular shell grown on its top [Fig.\ \ref{fig:SOI_Samples}(c)].
We describe the cross sections using the core radius $R_{\mathrm{c}}$ and shell radius $R_{\mathrm{s}}$, which represent the distances from the geometric center of the cross section to the vertices of the core and shell, respectively [Fig.\ \ref{fig:SOI_Samples}(a)]. Another parameter is the shell thickness $d$, shown in Figs.\ \ref{fig:SOI_Samples}(b) and \ref{fig:SOI_Samples}(c). In the Tri-Hex structure, the shell thickness $d$ is defined as the width of the shell region where the internal and external boundaries are parallel [Fig.\ \ref{fig:SOI_Samples}(c)].

In contrast to circular rings, electrons in polygonal rings are nonuniformly distributed along the \red{perimeter}. Specifically, low-energy electrons tend to localize in the corner regions, where they are attracted by effective quantum wells, as in the case of bent wires \cite{Sprung}. \red{The curved region of a bent wire behaves as a quantum well, whose depth and length are determined by the bend's angle and radius. A physically analogous effect arises in the corner regions of polygonal quantum rings. }
As a result, the probability distributions exhibit maxima in these areas (Fig.\ \ref{fig:L}, lower row).
For hexagonal cross sections, the maxima spread widely and overlap along the sides [Fig.\ \ref{fig:L}(a), lower panel], allowing the carriers to move around the cross section. However, in triangular and Tri-Hex structures, the electrons with the lowest energies are depleted from the sides [Figs.\ \ref{fig:L}(b) and \ref{fig:L}(c), lower row]. Here, the probability distributions form sharp peaks localized at the corners, leading to the formation of well-separated conductive channels. This means that a single shell may contain multiple one-dimensional wires, each confined to one of the sharp edges \cite{Ferrari09b}.
When the shell is confined internally and externally by the same polygon, higher transverse states are localized along the sides [Figs.\ \ref{fig:L}(a) and \ref{fig:L}(b), upper row] \cite{Ballester12,Sitek15}. In the Tri-Hex structure, the higher states remain localized at the corners, but the probability maxima form two peaks \red{in each corner} [Fig.\ \ref{fig:L}(c), upper row].

In Fig.\ \ref{fig:energy}, we compare the transverse energy states for the three cross sections shown in Fig.\ \ref{fig:SOI_Samples}. The energy states are either twofold or fourfold degenerate, due to spin only or spin and geometric symmetry, respectively, and are grouped according to their localization characteristics. The number of states in each group is twice the number of external vertices (Fig.\ \ref{fig:energy}). The dispersion of the lowest, corner-localized states decreases with the number of external vertices and side thickness \cite{Sitek15,Sitek16}. For sufficiently narrow triangular and Tri-Hex shells, the corner states become quasidegenerate [Figs.\ \ref{fig:energy}(b) and \ref{fig:energy}(c)]. If the shells were thicker, the lowest states would \red{visibly} split into a twofold degenerate ground state and a fourfold degenerate excited state \cite{Sitek16}. 

\red{The effective quantum wells associated with the corners deepen when the ratio between the effective linear size of the corner area and the thickness of the polygon sides increases, thus they become deeper when the number of corners decreases. As a result, the energy gap $\Delta$, which separates the corner states, increases with decreasing the number of corners and also side thickness \cite{Sitek15,Sitek16}.}

For the \red{8-nm-wide} hexagonal shell, the gap is larger than the energy differences within the corner-localized states but is comparable to the gaps between higher levels. In triangular and Tri-Hex shapes, the gap separating the six lowest states is significantly larger than other energy splittings and can even exceed the room-temperature energy. For the geometry and material parameters considered here, $\Delta_{\mathrm{Tri}}$ is slightly larger than $\Delta_{\mathrm{Tri\textrm{-}Hex}}$, but the case, where $\Delta_{\mathrm{Tri\textrm{-}Hex}}$ is considerably larger than $\Delta_{\mathrm{Tri}}$ is also possible \cite{Klausen20}. Moreover, a proper adjustment of the shell’s cross-section shape and material properties may result in an energy separation of the corner states exceeding 100 meV \cite{Sitek15, Klausen20}.
Although the energy gap $\Delta$ decreases with increasing shell thickness \cite{Sitek16}, for triangular and Tri-Hex structures it remains considerably large even for very wide shells. 

Thus, prismatic shells provide a \red{well-separated} subspace of corner-localized states that are expected to be robust against perturbations, such as weak disorder or cross-sectional symmetry breaking.

%%%---------------------------------------------------------------- 

\subsection{Nanowire states without SOI}

%%%%%%%%%%%%%%%%%%%%%%%%%%%%%%%%%%%%%%%%%%%%%%%%%%%%%%%%%
\begin{figure*}
	\centering
	\includegraphics[scale=0.67]{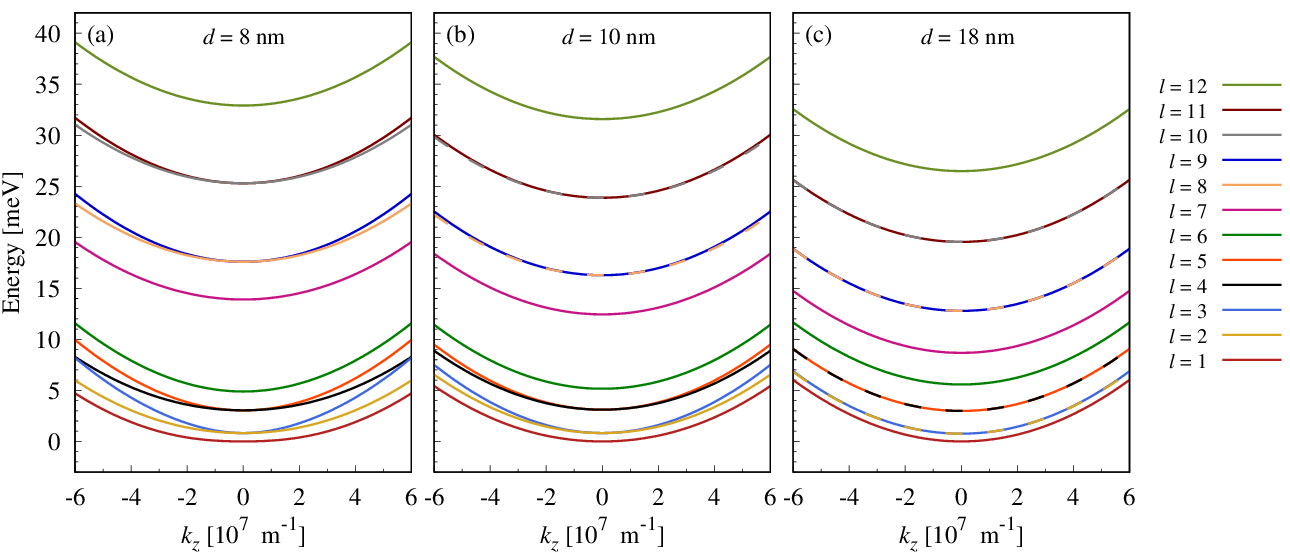}
	\caption{Energy dispersions for the corner ($l=1, \dots, 6$) and side ($l=7, \dots, 12$) states of hexagonal shells with varying side thicknesses ($d$), where $R_{\mathrm{c}}=40$ nm and $r=2$ nm. The energies are shifted with respect to the ground \red{state} energy at $k_z=0$. The line description defined on the right-hand side of the figure applies for all panels. \red{The label \(l\) refers to a pair of degenerate states.}  In (b) and (c) one of the overlapping curves in each pair (representing levels: 2, 4, 8, and 10) is shown as dashed, with the dashed curve color corresponding to the same energy level as the solid line in (a).}
	\label{fig:soi_hex_th}
%\end{figure*}
%%%%%%%%%%%%%%%%%%%%%%%%%%%%%%%%%%%%%%%%%%%%%%%%%%%%%%%%%
%%%%%%%%%%%%%%%%%%%%%%%%%%%%%%%%%%%%%%%%%%%%%%%%%%%%%%%%%
%%%--- Hexagon --- r = d --------------------------------
%\begin{figure*}
	\centering
	\includegraphics[scale=0.67]{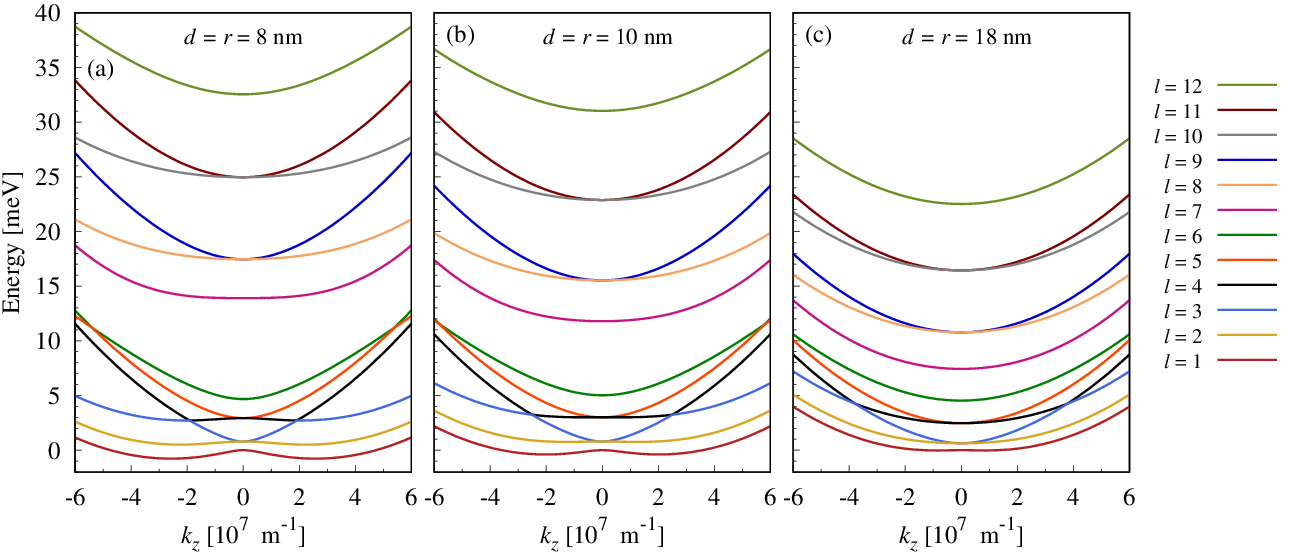}
	\caption{Energy dispersions for the corner ($l=1, \dots, 6$) and side ($l=7, \dots, 12$) states of hexagonal shells with varying side thicknesses ($d$) and for the intermixing lengths equal to the side thicknesses ($r=d$). Here, $R_{\mathrm{c}}=40$ nm. The energies are shifted with respect to the ground \red{state} energy at $k_z=0$. The line description defined on the right-hand side of the figure applies for all panels. \red{The label \(l\) refers to a pair of degenerate states.} }   
	\label{fig:soi_hex_r}
\end{figure*}
%%%%%%%%%%%%%%%%%%%%%%%%%%%%%%%%%%%%%%%%%%%%%%%%%%%%%%%%%

Now, we consider nanowires with infinite length. In the absence of SOI, the diagonalization of the Hamiltonian $\mathcal{H}_{cc}$ in Eq.\ (\ref{HSOI}) leads to parabolic energy dispersions, which reflect the degeneracy of the transverse states
\begin{eqnarray}
    E_{i}(k_z) = E_i + \frac{\hbar^2 k_z^2}{2m^{*}} \, ,
\end{eqnarray}
where $E_i$ represents the transverse energy states shown in Fig.\ \ref{fig:energy}, and $m^{*}$ is the effective mass defined in Eq.\ (\ref{m_eff_z}).

For the 8-nm-thick hexagonal cross section [Fig.\ \ref{fig:energy}(a)], the corner and side states lead to the formation of eight parabolic curves, as shown in Fig.\ \ref{fig:energy_kz}(a). Each group of states, corresponding to a particular localization, consists of 12 states arranged into four parabolas: The lowest parabola is always twofold degenerate (states $l=1$ and $l=5$), the next two curves are fourfold degenerate (states $l=2$, $3$, $6$ and $l=7$), and the fourth parabola is again twofold degenerate (states $l=4$ and $l=8$).

In the case of the 8-nm-wide triangular and Tri-Hex shells shown in Fig.\ \ref{fig:SOI_Samples}, the transverse corner states, visible in Figs.\ \ref{fig:energy}(b) and\ \ref{fig:energy}(c), are nearly degenerate. Consequently, for infinite-length wires, the corresponding energy levels form sixfold quasidegenerate parabolas, as shown in the lower panels of Figs.\ \ref{fig:energy_kz}(b) and \ref{fig:energy_kz}(c). 

In the case of the triangular shell, the six states localized in the facets [Fig.\ \ref{fig:L}(b)] can be described as follows: The first four transverse states, situated above the $\Delta_{\mathrm{Tri}}$ gap, form a fourfold degenerate level, which gives rise to the fourfold degenerate parabolic state ($l=3$) in Fig.\ \ref{fig:energy_kz}(b). The remaining two side-localized states are twofold degenerate, resulting in double parabolic states ($l=4$) in the 3D structure. In contrast, for the Tri-Hex shell, the states above the $\Delta_{\mathrm{Tri\textrm{-}Hex}}$ gap are also localized in the corners [Fig.\ \ref{fig:L}(c)]. Although these states are formally four- and twofold degenerate, the energy separation between these levels is so small [Fig.\ \ref{fig:energy}(c)] that they almost align and when considering infinite shells, lead to sixfold quasidegenerate parabolic states ($l=3$ and $l=4$) in Fig.\ \ref{fig:energy_kz}(c).

\subsection{Nanowire states with SOI}

The different band gaps and affinities of the core and shell materials create a potential difference at the core-shell interface. This potential difference induces an electric field, which, in turn, gives rise to SOI. The potential profile can be modeled during the growth process \cite{Fabian24} and can take a sharp, steplike shape or gradually decrease, i.e., can extend into the shell up to a distance $r$ (represented in dark blue in Fig.\ \ref{fig:SOI_Samples}). Additionally, SOI can be induced by an external electric field which changes the shape of the confining potential, and thus creates structural asymmetry.
The potential slope within the shell governs the Rashba coefficients which depend also on the energy gaps at the $\Gamma$ point of the shell [Eqs.\ (\ref{alpha_xy}) and\ (\ref{alpha0})].
The effect of SOI on the energy bands depends primary on the fraction of the cross-sectional area where the Rashba coefficients are nonzero and on the overlap of this region with the electron localization areas.

%%%----------------------------------------------------------------

\subsubsection{Hexagonal shell}

In Fig.\ \ref{fig:soi_hex_th}, we compare the energy dispersions of the corner and side states for symmetric hexagonal shells with varying side thicknesses. The shells surround identical cores ($R_{\mathrm{c}}=40$ nm) and the intermixing length ($r=2$ nm) is the same for each wire. 
The impact of SOI on the energy states is weak [Fig.\ \ref{fig:soi_hex_th}(a)], and it decreases with increasing energy, due to the shift of probability distribution towards the corner areas \red{\cite{Sitek19}, which} reduces the overlap between the wave functions and the region where the Rashba coefficients are nonzero (dark blue in Fig. \ref{fig:SOI_Samples}). 

\red{
In the presence of SOI, the energy dispersions remain even functions of $k_z$. SOI lifts only the fourfold degeneracies at $k_z\neq 0$, resulting in all states being twofold degenerate except at $k_z = 0$. To avoid overlapping lines and double labeling, in this section we use $l$ to denote pairs of degenerate states.
}
Although the $z$ spin component ($\sigma_z$) is no longer a good quantum number, time reversal symmetry holds, so all these states are doubly degenerate via Kramers' theorem. Nevertheless, the additional orbital degeneracy is still present at $k_z=0$, where SOI has no effect, and the energies are simply the transverse modes shown in Fig.\ \ref{fig:energy}(a) (before shifting the ground state to zero). This result is in disagreement with the cylindrical model of Rashba SOI, obtained by wrapping a planar electron gas on a cylinder \cite{Bringer11}, which lifts the orbital degeneracy at $k_z=0$ when imposed on a polygonal shell \cite{Manolescu17}.  Still, our present result is in agreement with the $\bm{k}\!\cdot\!\bm{p}$ calculations for the core component of a core-shell nanowire \cite{Wojcik18}.

%%%%%%%%%%%%%%%%%%%%%%%%%%%%%%%%%%%%%%%%%%%%%%%%%%%%%%%%%
\begin{figure}
\centering
     \includegraphics[scale=0.67]{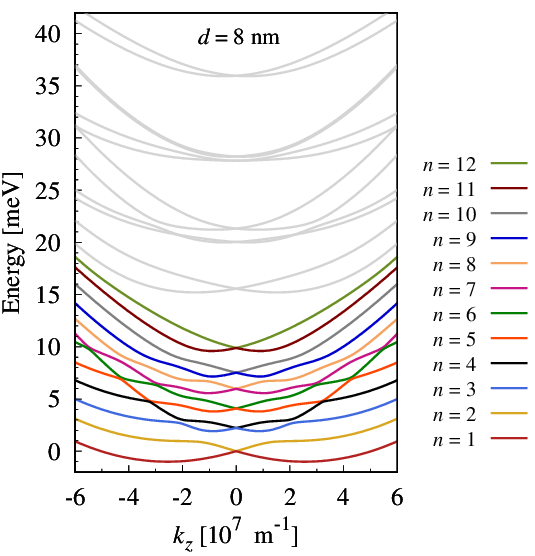}
\caption{Energy dispersions for the corner and side (gray lines) states of the 8-nm-wide hexagonal shell in the presence of an external electric field. Here, $R_{\mathrm{c}}=40$ nm, $r=8$ nm, $E=0.1$ mV/nm, and $\varphi=\pi/12$. The energies are shifted with respect to the ground \red{state} energy at $k_z=0$.}
\label{fig:soi_hex_E}
\end{figure}
%%%%%%%%%%%%%%%%%%%%%%%%%%%%%%%%%%%%%%%%%%%%%%%%%%%%%%%%%

As the shell thickness $d$ increases while the intermixing length $r$ is kept constant, the wave functions spread and the overlap with the region where the SOI coefficients are nonzero reduces.  At the same time, the distinction between corner and side states vanishes \red{and} the energy gap between these groups of states decreases. \red{Wide} shells lose the properties of a prismatic structure, approaching the behavior of cylindrical tubes \cite{Sitek16,Sitek19}.
The result is a gradual drop of the SOI effect, as seen in Figs.\ \ref{fig:soi_hex_th}(b) and \ref{fig:soi_hex_th}(c). In particular, for an 18-nm-wide shell [Fig.\ \ref{fig:soi_hex_th}(c)], the energy dispersions resemble the parabolic form \red{obtained} in the absence of SOI [Fig.\ \ref{fig:energy_kz}(a)].

The impact of SOI is significantly amplified when the core material penetrates deeper into the shell. To illustrate this statement, in Fig.\ \ref{fig:soi_hex_r} we repeat the examples shown in Fig.\ \ref{fig:soi_hex_th}, but for the intermixing length equal to the shell width ($r=d$). In this case, the band-offset potential gradient extends from the core-shell interface to the external shell boundary and the SOI coefficients are nonzero over the entire cross section, maximizing the area on which the spatial derivatives in Eq.\ (\ref{alpha_xy}) are nonzero \red{as well as their values}, and thus the SOI contribution to the Hamiltonian in Eq.\ (\ref{HSOI}). In addition, the tilted potential effectively narrows the shell by shifting the wave functions to the outer regions and modifying the electron distributions. As a result,
for all three shell widths studied the SOI effects are more pronounced than in Fig.\ \ref{fig:soi_hex_th}. 
In particular, the impact on the corner-localized states is significantly larger, the shift in the $\pm k_z$ directions is  considerable, such that the lowest corner states become non-monotonic functions for both $k_z<0$ and $k_z>0$, the energy splittings at finite $k_z$ are more significant, and the states $l=3$ and $l=4$ cross \red{at much smaller $k_z$}.

At larger $k_z$, the three lowest corner states change similarly with the wave vector along the growth direction. In contrast, the slopes of the three higher corner states ($l=4$, $5$, and $6$) decrease significantly with energy, leading to a reduction in the energy differences between them as $|k_z|$ increases, and thus to crossings. 
With the parameters used for Fig.\ \ref{fig:soi_hex_r}(a), the SOI effect on the side-localized states is smaller than on corner states, causing the energy dispersions to remain almost parabolic functions of $k_z$.
\green{This is due to the elongated shape of the side-localized maxima and their shift to regions where the extended band-offset potential is lower. }

%%%%%%%%%%%%%%%%%%%%%%%%%%%%%%%%%%%%%%%%%%%%%%%%%%%%%%%%%
\begin{figure*}
\centering
     \includegraphics[scale=0.67]{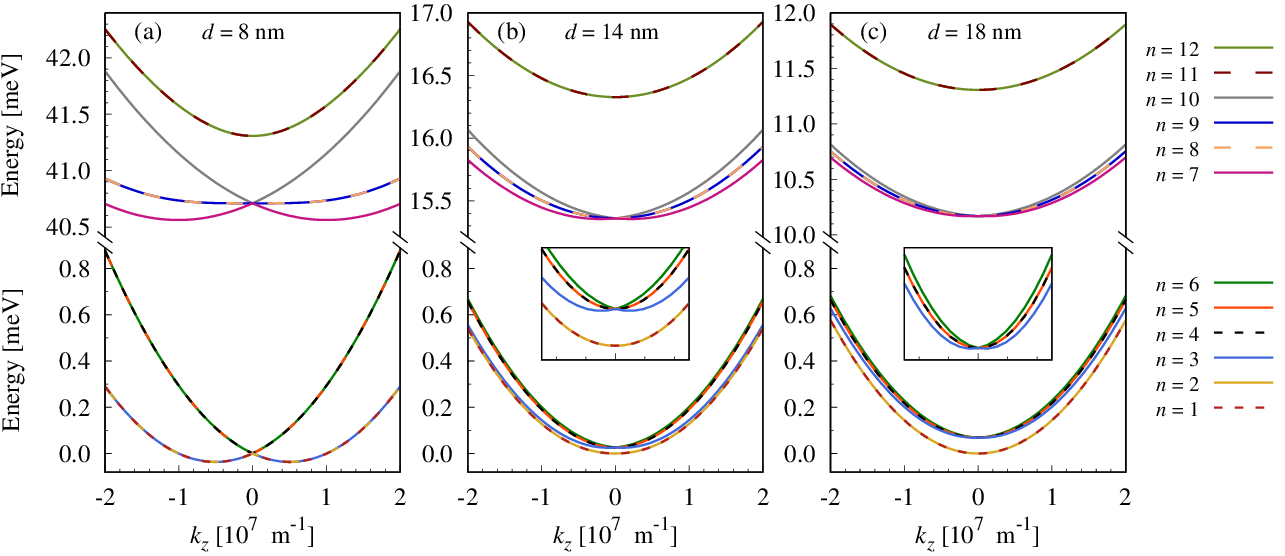}
\caption{Energy dispersions for the corner ($n=1, \dots, 6$) and side  ($n=7, \dots, 12$) states of triangular shells with varying side thicknesses ($d$). Here, $R_{\mathrm{c}}=40$ nm and $r=2$ nm. 
\red{The insets in (b) and (c) show a close-up view of the corner states around $k_z=0$. }
The energies are shifted with respect to the ground \red{state} energy at $k_z=0$. The line description defined on the right-hand side of the figure applies for all panels.}
\label{fig:soi_tri_th}
%\end{figure*}
%%%%%%%%%%%%%%%%%%%%%%%%%%%%%%%%%%%%%%%%%%%%%%%%%%%%%%%%%
%%%--- Triangle --- r = d -------------------------------
%\begin{figure*}
	\centering
	\includegraphics[scale=0.67]{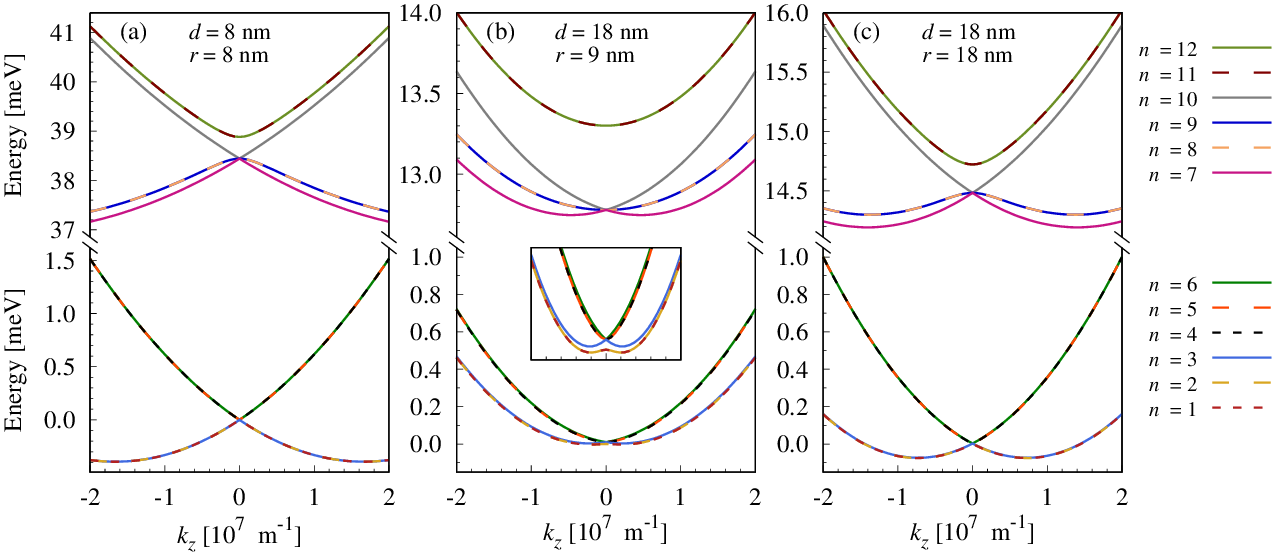}
\caption{Energy dispersions for the corner ($n=1, \dots, 6$) and side ($n=7, \dots, 12$) states of the 8 and 18-nm-wide triangular shells with different intermixing lengths ($r$). 
\red{The inset in (b) shows a close-up view of the corner states around $k_z=0$. }
Here, $R_{\mathrm{c}}=40$ nm. The energies are shifted with respect to the ground \red{state} energy at $k_z=0$. The line description defined on the right-hand side of the figure is valid for all panels.}
	\label{fig:soi_tri_r}
\end{figure*}
%%%%%%%%%%%%%%%%%%%%%%%%%%%%%%%%%%%%%%%%%%%%%%%%%%%%%%%%%

For all studied wires, the potential drops from a value of $V_{\mathrm{CBO}} = 656.6$ meV \cite{Vurgaftman01}. However, as the shell width increases while keeping $r = d$, the derivatives of the potential in Eqs.\ (\ref{alpha_xy}) decrease, leading to a reduced SOI contribution to the Hamiltonian in Eq.\ (\ref{HSOI}). Even so, that the impact of SOI on the energy states of the 10-nm-wide shell [Fig.\ \ref{fig:soi_hex_r}(b)] remains much stronger than when the potential penetrates only up to $r=2$ nm [Fig.\ \ref{fig:soi_hex_th}(b)]. Although the potential derivatives in Eqs.\ (\ref{alpha_xy}) are more than half smaller for the 18-nm-wide shell [Fig.\ \ref{fig:soi_hex_r}(c)] compared to the 8-nm shell, the effect of SOI on the energy states remains significant.

%%%----------------------------------------------------------------
%%%--- Hexagon --- electric field ---------------------------------

In Fig.\ \ref{fig:soi_hex_E}, we show the energy dispersions for the 8-nm-wide shell with $r = d$ [as in Fig.\ \ref{fig:soi_hex_r}(a)] in the presence of a perpendicular electric field rotated by $\varphi = \pi/12$ with respect to the $x$ axis (Fig.\ \ref{fig:SOI_Samples}).
The contribution to SOI caused by the electric field is much smaller than that due to the extended band-offset potential, but the impact of the field on the transverse energies and electron localization \cite{Sitek15}, and thus on the energy dispersions is considerable. The applied electric field lifts the fourfold degeneracies at $k_z = 0$ and the twofold degeneracies at finite $k_z$. As a result, the energies become twofold degenerate at $k_z = 0$, and nondegenerate elsewhere.
%
%\siggi{[Kramers degeneracy, again]}. 
The corner states exhibit a spectrum resembling six slightly different wires, with six pairs of parabolas symmetrically shifted in the $\pm k_z$ direction. However, in contrast to independent systems, the corner states do not cross but repel at the otherwise crossing points. This level repulsion depends on the overlap of the wave function \green{and is the indication of an interaction}.

The side states are more robust to an external electric field than the corner states \cite{Sitek15}. However, these states are still affected by the electric field and form complex distributions that can consist of multiple maxima, which is reflected in the energy dispersion (gray lines in Fig.\ \ref{fig:soi_hex_E}). The electric field induces energy gaps by lifting the fourfold degeneracies and increases the energy splittings, particularly for the corner states, but considerably reduces the energy gap between the corner and side states at $k_z = 0$, thereby decreasing the robustness of the corner states' subspace.

Realistic wires are never perfectly symmetric; Geometric distortions, such as variations in the thickness of the shell facets, break the wave function symmetries in a manner similar to the effect of a perpendicular electric field \cite{Sitek15, Sonner19}. As a result, the energy dispersion in the presence of an electric field closely resembles the spectrum of technologically accessible structures. Moreover, since an electrically induced asymmetry can cause a geometrically symmetric tube to behave as a nonsymmetric system, applying the electric field to geometrically distorted shells could, to some extent, restore the properties of symmetric structures.

%%%----------------------------------------------------------------

\subsubsection{Triangular shell}

%%%%%%%%%%%%%%%%%%%%%%%%%%%%%%%%%%%%%%%%%%%%%%%%%%%%%%%%%
\begin{figure}
	\centering
	\includegraphics[scale=0.67]{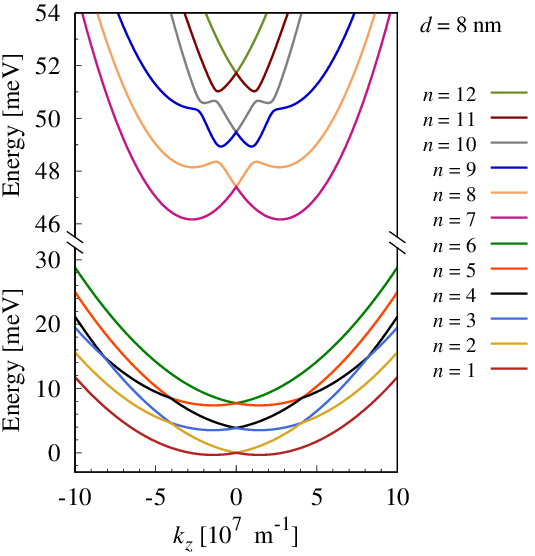}
	\caption{Energy dispersions for the corner ($n=1, \dots, 6$)  and side ($n=7, \dots, 12$) states of the 8-nm-wide triangular shell in the presence of an external electric field. Here, $R_{\mathrm{c}}=40$ nm, $r=d=8$ nm, $E=0.1$ mV/nm, $\varphi=\pi/6$. The energies are shifted with respect to the ground \red{state} energy at $k_z=0$.}
	\label{fig:soi_tri_E}
\end{figure}
%%%%%%%%%%%%%%%%%%%%%%%%%%%%%%%%%%%%%%%%%%%%%%%%%%%%%%%%%

The properties of quantum wires that originate from their polygonal cross sections, such as corner localization and the energy gap between the corner and side states, become more pronounced as the number of corners decreases. Thus these effects are the most significant for triangular structures \cite{Sitek15}. In Fig. \ref{fig:soi_tri_th}(a), we show the energy dispersion for an 8-nm-wide triangular shell. Although  the ground transverse state (twofold degenerate) and the first excited transverse state (fourfold degenerate)  are energetically resolved, they are very close to each other [Fig.\ \ref{fig:energy}(b)], and thus the corner-localized states can be considered a sixfold quasidegenerate level.
The corresponding probability distributions of the corner states consist of three well-separated, narrow peaks, so the SOI affects these states as it would affect three independent wires. The corner states [states $n=1$ to $n=6$ in Fig.\ \ref{fig:soi_tri_th}(a)] are thus arranged in three superimposed pairs, with the states shifted by wave vectors $\pm q = 0.44 \times 10^7$ m$^{-1}$, 
which corresponds to $\alpha =\frac{\hbar^2}{m^*}q \approx 15$ meV \hspace{-0.25 cm} nm when the approximation of shifted parabolas is used, i.e. 
\begin{eqnarray*}
  E = \frac{\hbar^2}{2 m^*} \left( k_z \pm q\right)^2 = \frac{\hbar^2 k_z^2}{2 m^*} \pm \alpha 
 % q 
 k_z + \frac{\hbar^2 q^2}{2 m^*}  \, . 
\end{eqnarray*}

The effect of SOI on the side states [from $n=7$ to $n=12$ in Fig.\ \ref{fig:soi_tri_th}(a)] looks qualitatively different. In particular, the lower state, fourfold degenerate at $k_z = 0$, splits into three states at finite $k_z$. The nondegenerate states ($n=7$ and $n=10$) arise from parabolas shifted by $\pm q\approx 0.88\times 10^7$ m$^{-1}$ ($\alpha\approx 30$ meV \hspace{-0.25 cm} nm), while the remaining two states ($n=8$ and $n=9$) form twofold quasiparabolic state with an effective mass that is larger than that predicted by Eq.\ (\ref{m_eff_z}). Similarly, the states $n=11$ and $n=12$ merge into one quasi-parabolic \red{state}, but with an effective mass smaller than that given in Eq.\ (\ref{m_eff_z}). 
Unlike a hexagonal cross section, the triangular cross section lacks inversion symmetry, and as a result, the degeneracies at $k_z \neq 0$ are partially lifted if the wave functions overlap.  
\green{Also, electrons with energies above the $\Delta_{\mathrm{Tri}}$ gap, i.e., above the corner states,} can occupy nearly the entire cross-sectional area, and thus the SOI affects these states differently than the \green{states locked at corners}.

\red{As in the case of hexagonal shells, the}  effect of SOI strongly decreases with increasing shell width \red{when $r$ is kept constant}
[Figs.\ \ref{fig:soi_tri_th}(b) and\ \ref{fig:soi_tri_th}(c)]. 
For wider shells, the corner states are arranged into two distinct levels at \red{$k_z =0$}, \green{the higher one splits into three states at finite $k_z$, thus behaving as the fourfold degenerate side-localized state.}  
Although the impact of SOI on the energies of the 18-nm triangular shell [Fig.\ \ref{fig:soi_tri_th}(c)] is weak, it is still stronger than in the case of the 18-nm-wide hexagonal wire [Fig.\ \ref{fig:soi_hex_th}(c)].

%%%----------------------------------------------------------------
%%%--- Triangle --- r = d -----------------------------------------

In Fig.\ \ref{fig:soi_tri_r} we show the results when the radial potential gradient extends over a wider range within the shell. For the 8-nm shell, the energy splitting of the corner states at $k_z=2\times 10^7$ $\mathrm{m}^{-1}$ increases by approximately a factor of three [Figs.\ \ref{fig:soi_tri_th}(a) vs.\ \ref{fig:soi_tri_r}(a)]. Moreover, the side states become more similar to the corner states, and at finite $k_z$, they are arranged into two well-separated groups consisting of three closely spaced states [Fig.\ \ref{fig:soi_tri_r}(a)].
Similarly to hexagonal wires, the increased SOI effect is related to the enlarged region with potential gradient and effectively increased confinement.
Even if the \red{P atoms} penetrate only up to half the width of the 18-nm-wide shell [Fig.\ \ref{fig:soi_tri_r}(b)], the effects of SOI increase significantly with respect to the case when 
$r=2$ nm \red{[Fig.\ \ref{fig:soi_tri_th}(c)]}, and the energy dispersions resemble those of thinner shells [Fig.\ \ref{fig:soi_tri_th}(a)]. For $r=d$, the energy dispersions of the 8-nm-wide [Fig.\ \ref{fig:soi_tri_r}(a)] and 18-nm-thick [Fig.\ \ref{fig:soi_tri_r}(c)] shells are qualitatively similar, though quantitatively weaker in the latter case.

%%%----------------------------------------------------------------
%%%--- Triangle --- electric field --------------------------------

If a triangular wire is exposed to a perpendicular electric field rotated by an angle
$\varphi=\pi/6$ with respect to the $x$ axis, the six transverse corner states are rearranged into three twofold degenerate levels. Each level corresponds to a probability distribution with a sharp maximum localized in a specific corner area \cite{Sitek15}. Since the corner-localized maxima of the 8-nm-wide shell are separated by large areas from which electrons are depleted [Fig.\ \ref{fig:L}(b)], the eigenstates associated with the three lowest energies do not overlap in the presence of the electric field. Therefore, in the low-energy regime, the triangular shell behaves as three independent wires with different ground energies. 
%
%%%\siggi{[does the SOI strength change?]} \green{For the corner states and the lowest side state no, for the 2 higher side states - yes.} 
%
This behavior is illustrated in Fig.\ \ref{fig:soi_tri_E}, where the six lowest states form three pairs of parabolas shifted in the $\pm k_z$ direction which cross at $k_z=0$ as well as at finite values of $k_z$, \green{which confirms the absence of interaction}. 
%
%%%\siggi{[try to use symmetry arguments]} \green{Why? What for?}
%

The side-localized wave functions, which form elongated maxima [Fig.\ \ref{fig:L}(b)], are much more robust to the electric field than the corner states \cite{Sitek15}, and thus they overlap. Consequently, in the presence of SOI, the side states cross only at $k_z=0$
and repel each other for $k_z \neq 0$ when they come close \green{which indicates the presence of interaction}.

One special feature of the triangular structures is the large energy gap that protects the corner states from interactions with other states even in the presence of an electric field. This gap also allows for the manipulation of the corner state subspace; for example, by rotating the electric field, one can tune the energy splittings between the corner states at $k_z=0$  \cite{Sitek17ICTON}.

%%%----------------------------------------------------------------

\subsubsection{Tri-Hex shell}

%%%%%%%%%%%%%%%%%%%%%%%%%%%%%%%%%%%%%%%%%%%%%%%%%%%%%%%%%
\begin{figure}
\centering
     \includegraphics[scale=0.67]{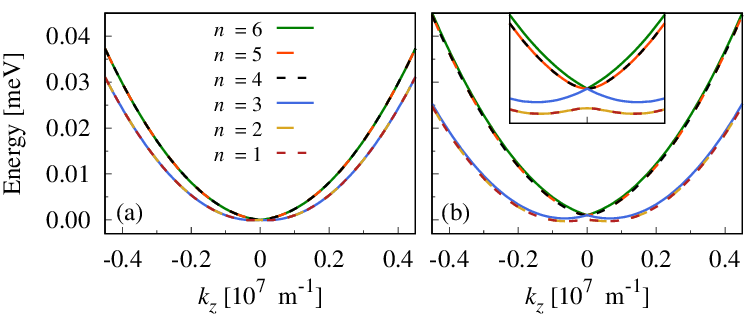}
\caption{Energy dispersions for the corner states of the 8-nm-wide Tri-Hex shells grown on different cores: (a) $R_{\mathrm{c}}=40$ nm, (b) $R_{\mathrm{c}}=20$ nm. 
\red{The inset in (b) shows a close-up view of the states around $k_z=0$. }
The energies are shifted with respect to the ground \red{state} energy at $k_z=0$.}
\label{fig:soi_tri_hex_th}
\end{figure}
%%%%%%%%%%%%%%%%%%%%%%%%%%%%%%%%%%%%%%%%%%%%%%%%%%%%%%%%%

The transverse energies [Fig.\ \ref{fig:energy}(c)] and the low-energy electron localization [Fig.\ \ref{fig:L}(c)] of the Tri-Hex shells resemble those of the triangular ones. Since the $\Delta_{\mathrm{Tri\textrm{-}Hex}}$ gap can be comparable to or even larger than the \green {corresponding value for} triangular wires \green{($\Delta_{\mathrm{Tri}}$)} \cite{Klausen20}, and the states above $\Delta_{\mathrm{Tri\textrm{-}Hex}}$ are localized in the corner regions, we focus  \green{on the six lowest states localized in the corners.} 

The corner areas of Tri-Hex structures are much larger than those of hexagonal and triangular shells grown on cores with the same radius (Fig.\ \ref{fig:SOI_Samples}). When $r = 2$ nm, the area where the SOI coefficients [Eq.\ (\ref{alpha_xy})] are nonzero is the same as in the case of hexagonal wires. However, this area constitutes a very small fraction of the Tri-Hex cross section, so the impact of SOI is minimal [Fig.\ \ref{fig:soi_tri_hex_th}(a)]. 
\green{Further, we increased this fraction by reducing the core radius by half while keeping $r=2$ nm and $d = 8$ nm, which resulted in the amplification of the SOI [Fig.\ \ref{fig:soi_tri_hex_th}(b)]. Moreover, the energy splitting between the ground state and the first excited state at $k_z=0$ increased [inset to Fig.\ \ref{fig:soi_tri_hex_th}(b)], along with the ratio $d/R_{\mathrm{s}}$.}

%%%----------------------------------------------------------------
%%%--- Tri-Hex --- r = d ------------------------------------------

%%%%%%%%%%%%%%%%%%%%%%%%%%%%%%%%%%%%%%%%%%%%%%%%%%%%%%%%%
\begin{figure}
\centering
     \includegraphics[scale=0.67]{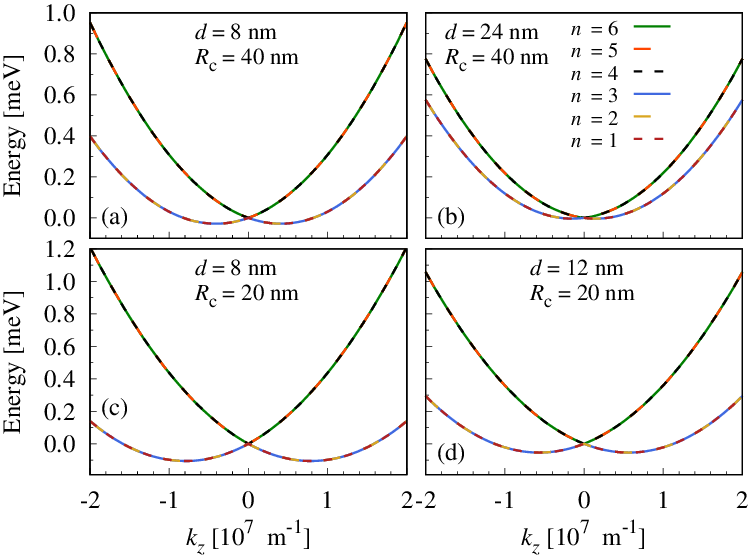}
\caption{Energy dispersions for the corner states of different Tri-Hex shells, with the intermixing length equal to 75\% of the corner area's height. The energies are shifted with respect to the ground \red{state} energy at $k_z=0$.}
\label{fig:soi_tri_hex_r}
\end{figure}
%%%%%%%%%%%%%%%%%%%%%%%%%%%%%%%%%%%%%%%%%%%%%%%%%%%%%%%%%
%%%%%%%%%%%%%%%%%%%%%%%%%%%%%%%%%%%%%%%%%%%%%%%%%%%%%%%%%
\begin{figure}
\centering
     \includegraphics[scale=0.67]{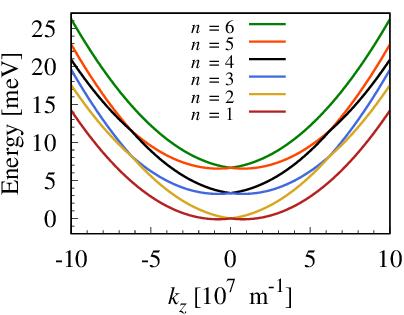}
\caption{Energy dispersions for the corner states of the 8-nm-wide Tri-Hex shell in the presence of an external electric field. Here, $R_{\mathrm{c}}=20$ nm, $r=d=8$ nm, $E=0.1$ mV/nm, $\varphi=\pi/6$. The energies are shifted with respect to the ground \red{state} energy at $k_z=0$.}
\label{fig:soi_Tri-Hex_E}
\end{figure}
%%%%%%%%%%%%%%%%%%%%%%%%%%%%%%%%%%%%%%%%%%%%%%%%%%%%%%%%%

As in the case of hexagonal and triangular shells, we increase the parameter $r$, representing the intermixing length of the \green{core and shell materials}, to 75\% of the distance between the core-shell interface and the shell external vertices. The enhancement due to the extended band-offset potential is considerably stronger than for the triangular wires. The energy dispersions for the 8-nm-wide shell [Figs.\ \ref{fig:soi_tri_hex_r}(a) and\ \ref{fig:soi_tri_hex_r}(c)], grown either on a core inscribed in a 40- or 20-nm circle, \red{indicate} that both structures behave as three independent and identical wires. 
\green{The absence of interaction is also confirmed by the crossings of energy states at finite $k_z$ in the presence of a perpendicular electric field,} \red {when the three effective wires are no longer identical (Fig.\ \ref{fig:soi_Tri-Hex_E}). }
Moreover, the strength of SOI is of the order of that obtained for the triangular shells, i.e., $\alpha$ varies between 4 and 28 $\mathrm{meV} \hspace{0.03 cm}\mathrm{nm}$. 

Interestingly, when the core material penetrates deep into the shell, the effects due to SOI are relatively robust to the increase in shell width. If $R_{\mathrm{c}} = 40$ nm, then increasing $d$ from 8 to 24 nm reduces the impact of SOI, but it remains significant [Fig.\ \ref{fig:soi_tri_hex_r}(b)]. For the smaller shell, increasing $d$ by 50\% only weakly influences the energy dispersions [Fig.\ \ref{fig:soi_tri_hex_r}(d)]. Due to its geometry, an increase in shell width results in a smaller relative enlargement of the Tri-Hex corner areas compared to triangular and hexagonal shell structures. As a result, the change in the effect of SOI is also smaller.

%%%----------------------------------------------------------------

\section{\label{sec:cons}Conclusions}

In summary, we studied the impact of the SOI due to the extended band-offset potential and an external electric field on the energies of hexagonal, triangular, and Tri-Hex tubular wires, which are the outer layers of core-shell nanowires. 
\red{Our results are derived with the $\bm{k}\!\cdot\! \bm{p}$ method for the shell and are compatible to calculations performed by other authors \cite{Wojcik19,Woods19,Campos18} for the core.  The presence of SOI in the shell has been experimentally observed, for example, in a recent study of magnetoconductance oscillations \cite{Basaric25}, and to the best of our knowledge, band-structure-based models of SOI in prismatic shells are missing or are rare in the literature.}

We \red{show} that, if the SOI is caused only by the conduction-band offset along the heterojunction, then the effect of SOI on the energies of hexagonal shells is qualitatively different from the impact on the spectra of triangular and Tri-Hex structures.
For all three geometries studied \red{(hexagonal, triangular, and combined)}, the SOI due to the band-offset potential lifts only the fourfold degeneracies at finite $k_z$, but in the case of hexagonal tubes, it splits the states into two twofold degenerate states such that the states of hexagonal shells become pairwise degenerate (though spin is no longer conserved). In contrast, the fourfold degenerate levels of triangular and Tri-Hex tubes split into three states: The lowest and highest states are nondegenerate, while the middle state is twofold degenerate.

Qualitatively, the SOI due to the extended band-offset potential affects the corner- and side-localized states in the same way, but the impact on the lower-energy states is stronger. 
If the intermixing length is kept short and constant while the shell thickness increases, i.e., as clean shells are grown, the impact of SOI on the energy states decreases together with the cross-sectional fraction, where the SOI coefficients are nonzero. However, when the shell thickness increases the SOI has a much stronger effect on the energies of the triangular tubes compared to the hexagonal ones.

The SOI in polygonal tubes can be enhanced or controlled through engineering of the band-offset potential. Specifically, if a tilted potential that penetrates deep into the shell is created, i.e., an extended band-offset potential is present, the SOI coefficients become nonzero over a significant fraction of the cross-sectional area. This leads to a considerable enhancement of SOI which, according to our results, for triangular and hexagonal wires is maximized for $r=d$, although for Tri-Hex structures the strongest effects occur for $r<d$.  Such SOI engineering
is advantageous, for example, in the context of Majorana states. Additionally, the extended band-offset potential increases the confinement, thereby amplifying characteristics associated with polygonal cross sections, such as electron accumulation along the sharp edges and the energetic separation of the lowest corner states.

The energy gap separating the corner states of the triangular and Tri-Hex wires from the higher states can significantly exceed the room-temperature energy \cite{Sitek15,Klausen20}. Moreover, these structures maintain the unique properties associated with their polygonal geometry over a broad range of shell thicknesses.
 
From a prospective application standpoint, the properties of triangular wires are particularly interesting, as  triangular wires can act as independent wires at low energies and as interacting ones at higher energies. Although both, the corner and side states, of triangular shells are significantly affected by SOI, the impact on the two groups can differ, which also opens new possibilities for applications.

\vspace{0.5 cm}

The data supporting the figures of this paper are available at the Zenodo repository \cite{Zenodo_SOI}.

%%%----------------------------------------------------------------

\begin{acknowledgments}
%\red{Andrei and Siggi, check the number.}	Done
This research was supported by the Icelandic Research Fund, Grant No. 195943.
We are grateful to Anna Musiał and Krzysztof Gawarecki for fruitful discussions.
\end{acknowledgments}

%%%----------------------------------------------------------------

%\bibliographystyle{apsrev4-2}
%\bibliography{core_shell}

\begin{thebibliography}{60}%
\makeatletter
\providecommand \@ifxundefined [1]{%
 \@ifx{#1\undefined}
}%
\providecommand \@ifnum [1]{%
 \ifnum #1\expandafter \@firstoftwo
 \else \expandafter \@secondoftwo
 \fi
}%
\providecommand \@ifx [1]{%
 \ifx #1\expandafter \@firstoftwo
 \else \expandafter \@secondoftwo
 \fi
}%
\providecommand \natexlab [1]{#1}%
\providecommand \enquote  [1]{``#1''}%
\providecommand \bibnamefont  [1]{#1}%
\providecommand \bibfnamefont [1]{#1}%
\providecommand \citenamefont [1]{#1}%
\providecommand \href@noop [0]{\@secondoftwo}%
\providecommand \href [0]{\begingroup \@sanitize@url \@href}%
\providecommand \@href[1]{\@@startlink{#1}\@@href}%
\providecommand \@@href[1]{\endgroup#1\@@endlink}%
\providecommand \@sanitize@url [0]{\catcode `\\12\catcode `\$12\catcode
  `\&12\catcode `\#12\catcode `\^12\catcode `\_12\catcode `\%12\relax}%
\providecommand \@@startlink[1]{}%
\providecommand \@@endlink[0]{}%
\providecommand \url  [0]{\begingroup\@sanitize@url \@url }%
\providecommand \@url [1]{\endgroup\@href {#1}{\urlprefix }}%
\providecommand \urlprefix  [0]{URL }%
\providecommand \Eprint [0]{\href }%
\providecommand \doibase [0]{https://doi.org/}%
\providecommand \selectlanguage [0]{\@gobble}%
\providecommand \bibinfo  [0]{\@secondoftwo}%
\providecommand \bibfield  [0]{\@secondoftwo}%
\providecommand \translation [1]{[#1]}%
\providecommand \BibitemOpen [0]{}%
\providecommand \bibitemStop [0]{}%
\providecommand \bibitemNoStop [0]{.\EOS\space}%
\providecommand \EOS [0]{\spacefactor3000\relax}%
\providecommand \BibitemShut  [1]{\csname bibitem#1\endcsname}%
\let\auto@bib@innerbib\@empty
%</preamble>
\bibitem [{\citenamefont {Nadj-Perge}\ \emph {et~al.}(2010)\citenamefont
  {Nadj-Perge}, \citenamefont {Frolov}, \citenamefont {Bakkers},\ and\
  \citenamefont {Kouwenhoven}}]{Nadj_Perge10}%
  \BibitemOpen
  \bibfield  {author} {\bibinfo {author} {\bibfnamefont {S.}~\bibnamefont
  {Nadj-Perge}}, \bibinfo {author} {\bibfnamefont {S.~M.}\ \bibnamefont
  {Frolov}}, \bibinfo {author} {\bibfnamefont {E.~P. A.~M.}\ \bibnamefont
  {Bakkers}},\ and\ \bibinfo {author} {\bibfnamefont {L.~P.}\ \bibnamefont
  {Kouwenhoven}},\ }\bibfield  {title} {\bibinfo {title} {{Spin-orbit qubit
  in a semiconductor nanowire}},\ }\href {https://doi.org/10.1038/nature09682}
  {\bibfield  {journal} {\bibinfo  {journal} {Nature}\ }\textbf {\bibinfo
  {volume} {468}},\ \bibinfo {pages} {1084} (\bibinfo {year}
  {2010})}\BibitemShut {NoStop}%
\bibitem [{\citenamefont {van~den Berg}\ \emph {et~al.}(2013)\citenamefont
  {van~den Berg}, \citenamefont {Nadj-Perge}, \citenamefont {Pribiag},
  \citenamefont {Plissard}, \citenamefont {Bakkers}, \citenamefont {Frolov},\
  and\ \citenamefont {Kouwenhoven}}]{van_den_Berg13}%
  \BibitemOpen
  \bibfield  {author} {\bibinfo {author} {\bibfnamefont {J.~W.~G.}\
  \bibnamefont {van~den Berg}}, \bibinfo {author} {\bibfnamefont
  {S.}~\bibnamefont {Nadj-Perge}}, \bibinfo {author} {\bibfnamefont {V.~S.}\
  \bibnamefont {Pribiag}}, \bibinfo {author} {\bibfnamefont {S.~R.}\
  \bibnamefont {Plissard}}, \bibinfo {author} {\bibfnamefont {E.~P. A.~M.}\
  \bibnamefont {Bakkers}}, \bibinfo {author} {\bibfnamefont {S.~M.}\
  \bibnamefont {Frolov}},\ and\ \bibinfo {author} {\bibfnamefont {L.~P.}\
  \bibnamefont {Kouwenhoven}},\ }\bibfield  {title} {\bibinfo {title} {Fast
  spin-orbit qubit in an indium antimonide nanowire},\ }\href
  {https://doi.org/10.1103/PhysRevLett.110.066806} {\bibfield  {journal}
  {\bibinfo  {journal} {Phys. Rev. Lett.}\ }\textbf {\bibinfo {volume} {110}},\
  \bibinfo {pages} {066806} (\bibinfo {year} {2013})}\BibitemShut {NoStop}%
\bibitem [{\citenamefont {Koo}\ \emph {et~al.}(2009)\citenamefont {Koo},
  \citenamefont {Kwon}, \citenamefont {Eom}, \citenamefont {Chang},
  \citenamefont {Han},\ and\ \citenamefont {Johnson}}]{Koo09}%
  \BibitemOpen
  \bibfield  {author} {\bibinfo {author} {\bibfnamefont {H.~C.}\ \bibnamefont
  {Koo}}, \bibinfo {author} {\bibfnamefont {J.~H.}\ \bibnamefont {Kwon}},
  \bibinfo {author} {\bibfnamefont {J.}~\bibnamefont {Eom}}, \bibinfo {author}
  {\bibfnamefont {J.}~\bibnamefont {Chang}}, \bibinfo {author} {\bibfnamefont
  {S.~H.}\ \bibnamefont {Han}},\ and\ \bibinfo {author} {\bibfnamefont
  {M.}~\bibnamefont {Johnson}},\ }\bibfield  {title} {\bibinfo {title} {Control
  of spin precession in a spin-injected field effect transistor},\ } {\bibfield  {journal} {\bibinfo
  {journal} {Science}\ }\textbf {\bibinfo {volume} {325}},\ \bibinfo {pages}
  {1515} (\bibinfo {year} {2009})} \BibitemShut
  {NoStop}%
\bibitem [{\citenamefont {Wójcik}\ \emph {et~al.}(2014)\citenamefont
  {Wójcik}, \citenamefont {Adamowski}, \citenamefont {Spisak},\ and\
  \citenamefont {Wołoszyn}}]{Wojcik14}%
  \BibitemOpen
  \bibfield  {author} {\bibinfo {author} {\bibfnamefont {P.}~\bibnamefont
  {Wójcik}}, \bibinfo {author} {\bibfnamefont {J.}~\bibnamefont {Adamowski}},
  \bibinfo {author} {\bibfnamefont {B.~J.}\ \bibnamefont {Spisak}},\ and\
  \bibinfo {author} {\bibfnamefont {M.}~\bibnamefont {Wołoszyn}},\ }\bibfield
  {title} {\bibinfo {title} {{Spin transistor operation driven by the Rashba
  spin-orbit coupling in the gated nanowire}},\ } {\bibfield  {journal} {\bibinfo
  {journal} {Journal of Applied Physics}\ }\textbf {\bibinfo {volume} {115}},\
  \bibinfo {pages} {104310} (\bibinfo {year} {2014})}
  \BibitemShut {NoStop}%
\bibitem [{\citenamefont {Ngo}\ \emph {et~al.}(2010)\citenamefont {Ngo},
  \citenamefont {Debray},\ and\ \citenamefont {Ulloa}}]{Ngo10}%
  \BibitemOpen
  \bibfield  {author} {\bibinfo {author} {\bibfnamefont {A.~T.}\ \bibnamefont
  {Ngo}}, \bibinfo {author} {\bibfnamefont {P.}~\bibnamefont {Debray}},\ and\
  \bibinfo {author} {\bibfnamefont {S.~E.}\ \bibnamefont {Ulloa}},\ }\bibfield
  {title} {\bibinfo {title} {Lateral spin-orbit interaction and spin
  polarization in quantum point contacts},\ }\href
  {https://doi.org/10.1103/PhysRevB.81.115328} {\bibfield  {journal} {\bibinfo
  {journal} {Phys. Rev. B}\ }\textbf {\bibinfo {volume} {81}},\ \bibinfo
  {pages} {115328} (\bibinfo {year} {2010})}\BibitemShut {NoStop}%
\bibitem [{\citenamefont {Kohda}\ \emph {et~al.}(2013)\citenamefont {Kohda},
  \citenamefont {Nakamura}, \citenamefont {Nishihara}, \citenamefont
  {Kobayashi}, \citenamefont {Ono}, \citenamefont {Ohe}, \citenamefont
  {Tokura}, \citenamefont {Mineno},\ and\ \citenamefont {Nitta}}]{Kohda13}%
  \BibitemOpen
  \bibfield  {author} {\bibinfo {author} {\bibfnamefont {M.}~\bibnamefont
  {Kohda}}, \bibinfo {author} {\bibfnamefont {S.}~\bibnamefont {Nakamura}},
  \bibinfo {author} {\bibfnamefont {Y.}~\bibnamefont {Nishihara}}, \bibinfo
  {author} {\bibfnamefont {K.}~\bibnamefont {Kobayashi}}, \bibinfo {author}
  {\bibfnamefont {T.}~\bibnamefont {Ono}}, \bibinfo {author} {\bibfnamefont
  {J.-i.}\ \bibnamefont {Ohe}}, \bibinfo {author} {\bibfnamefont
  {Y.}~\bibnamefont {Tokura}}, \bibinfo {author} {\bibfnamefont
  {T.}~\bibnamefont {Mineno}},\ and\ \bibinfo {author} {\bibfnamefont
  {J.}~\bibnamefont {Nitta}},\ }\bibfield  {title} {\bibinfo {title}
  {Spin-orbit induced electronic spin separation in semiconductor
  nanostructures},\ }\href {https://doi.org/10.1038/ncomms2080} {\bibfield
  {journal} {\bibinfo  {journal} {Nat. Commun.}\ }\textbf {\bibinfo {volume}
  {3}},\ \bibinfo {pages} {1082} (\bibinfo {year} {2013})}\BibitemShut
  {NoStop}%
\bibitem [{\citenamefont {Wójcik}\ and\ \citenamefont
  {Adamowski}(20)}]{Wojcik17}%
  \BibitemOpen
  \bibfield  {author} {\bibinfo {author} {\bibfnamefont {P.}~\bibnamefont
  {Wójcik}}\ and\ \bibinfo {author} {\bibfnamefont {J.}~\bibnamefont
  {Adamowski}},\ }\bibfield  {title} {\bibinfo {title} {Spin filtering effect
  generated by the inter-subband spin-orbit coupling in the bilayer nanowire
  with the quantum point contact},\ }\href {https://doi.org/10.1038/srep45346}
  {\bibfield  {journal} {\bibinfo  {journal} {Sci. Rep.}\ }\textbf {\bibinfo
  {volume} {7}},\ \bibinfo {pages} {45346} (\bibinfo {year} {20})}\BibitemShut
  {NoStop}%
\bibitem [{\citenamefont {Lutchyn}\ \emph {et~al.}(2010)\citenamefont
  {Lutchyn}, \citenamefont {Sau},\ and\ \citenamefont {Das~Sarma}}]{Lutchyn10}%
  \BibitemOpen
  \bibfield  {author} {\bibinfo {author} {\bibfnamefont {R.~M.}\ \bibnamefont
  {Lutchyn}}, \bibinfo {author} {\bibfnamefont {J.~D.}\ \bibnamefont {Sau}},\
  and\ \bibinfo {author} {\bibfnamefont {S.}~\bibnamefont {Das~Sarma}},\
  }\bibfield  {title} {\bibinfo {title} {Majorana fermions and a topological
  phase transition in semiconductor-superconductor heterostructures},\ }\href
  {https://doi.org/10.1103/PhysRevLett.105.077001} {\bibfield  {journal}
  {\bibinfo  {journal} {Phys. Rev. Lett.}\ }\textbf {\bibinfo {volume} {105}},\
  \bibinfo {pages} {077001} (\bibinfo {year} {2010})}\BibitemShut {NoStop}%
\bibitem [{\citenamefont {Oreg}\ \emph {et~al.}(2010)\citenamefont {Oreg},
  \citenamefont {Refael},\ and\ \citenamefont {von Oppen}}]{Oreg10}%
  \BibitemOpen
  \bibfield  {author} {\bibinfo {author} {\bibfnamefont {Y.}~\bibnamefont
  {Oreg}}, \bibinfo {author} {\bibfnamefont {G.}~\bibnamefont {Refael}},\ and\
  \bibinfo {author} {\bibfnamefont {F.}~\bibnamefont {von Oppen}},\ }\bibfield
  {title} {\bibinfo {title} {Helical liquids and Majorana bound states in
  quantum wires},\ }\href {https://doi.org/10.1103/PhysRevLett.105.177002}
  {\bibfield  {journal} {\bibinfo  {journal} {Phys. Rev. Lett.}\ }\textbf
  {\bibinfo {volume} {105}},\ \bibinfo {pages} {177002} (\bibinfo {year}
  {2010})}\BibitemShut {NoStop}%
\bibitem [{\citenamefont {Stanescu}(2017)}]{Stanescu17}%
  \BibitemOpen
  \bibfield  {author} {\bibinfo {author} {\bibfnamefont {T.~D.}\ \bibnamefont
  {Stanescu}},\ }\href@noop {} {\emph {\bibinfo {title} {Introduction to
  Topological Quantum Matter and Quantum Computation}}}\ (\bibinfo  {publisher}
  {CRC Press},\ \bibinfo {address} {Oxford, U.K.},\ \bibinfo {year}
  {2017})\BibitemShut {NoStop}%
\bibitem [{\citenamefont {Laubscher}\ and\ \citenamefont
  {Klinovaja}(2021)}]{Laubscher21}%
  \BibitemOpen
  \bibfield  {author} {\bibinfo {author} {\bibfnamefont {K.}~\bibnamefont
  {Laubscher}}\ and\ \bibinfo {author} {\bibfnamefont {J.}~\bibnamefont
  {Klinovaja}},\ }\bibfield  {title} {\bibinfo {title} {{Majorana bound states
  in semiconducting nanostructures}},\ } {\bibfield  {journal} {\bibinfo
  {journal} {Journal of Applied Physics}\ }\textbf {\bibinfo {volume} {130}},\
  \bibinfo {pages} {081101} (\bibinfo {year} {2021})}
  \BibitemShut {NoStop}%
\bibitem [{\citenamefont {Manolescu}\ \emph {et~al.}(2017)\citenamefont
  {Manolescu}, \citenamefont {Sitek}, \citenamefont {Osca}, \citenamefont
  {Serra}, \citenamefont {Gudmundsson},\ and\ \citenamefont
  {Stanescu}}]{Manolescu17}%
  \BibitemOpen
  \bibfield  {author} {\bibinfo {author} {\bibfnamefont {A.}~\bibnamefont
  {Manolescu}}, \bibinfo {author} {\bibfnamefont {A.}~\bibnamefont {Sitek}},
  \bibinfo {author} {\bibfnamefont {J.}~\bibnamefont {Osca}}, \bibinfo {author}
  {\bibfnamefont {L.}~\bibnamefont {Serra}}, \bibinfo {author} {\bibfnamefont
  {V.}~\bibnamefont {Gudmundsson}},\ and\ \bibinfo {author} {\bibfnamefont
  {T.~D.}\ \bibnamefont {Stanescu}},\ }\bibfield  {title} {\bibinfo {title}
  {Majorana states in prismatic core-shell nanowires},\ }\href@noop {}
  {\bibfield  {journal} {\bibinfo  {journal} {Phys. Rev. B}\ }\textbf {\bibinfo
  {volume} {96}},\ \bibinfo {pages} {125435} (\bibinfo {year}
  {2017})}\BibitemShut {NoStop}%
\bibitem [{\citenamefont {Stanescu}\ \emph {et~al.}(2018)\citenamefont
  {Stanescu}, \citenamefont {Sitek},\ and\ \citenamefont
  {Manolescu}}]{Stanescu18}%
  \BibitemOpen
  \bibfield  {author} {\bibinfo {author} {\bibfnamefont {T.~D.}\ \bibnamefont
  {Stanescu}}, \bibinfo {author} {\bibfnamefont {A.}~\bibnamefont {Sitek}},\
  and\ \bibinfo {author} {\bibfnamefont {A.}~\bibnamefont {Manolescu}},\
  }\bibfield  {title} {\bibinfo {title} {Robust topological phase in
  proximitized core-shell nanowires coupled to multiple superconductors},\
  }\href@noop {} {\bibfield  {journal} {\bibinfo  {journal} {Beilstein J.
  Nanotechnol.}\ }\textbf {\bibinfo {volume} {9}},\ \bibinfo {pages} {1512}
  (\bibinfo {year} {2018})}\BibitemShut {NoStop}%
\bibitem [{\citenamefont {Kim}\ and\ \citenamefont {No}(2017)}]{Kim17}%
  \BibitemOpen
  \bibfield  {author} {\bibinfo {author} {\bibfnamefont {K.-H.}\ \bibnamefont
  {Kim}}\ and\ \bibinfo {author} {\bibfnamefont {Y.-S.}\ \bibnamefont {No}},\
  }\bibfield  {title} {\bibinfo {title} {Subwavelength core/shell cylindrical
  nanostructures for novel plasmonic and metamaterial devices},\ }\href
  {https://doi.org/10.1186/s40580-017-0128-8} {\bibfield  {journal} {\bibinfo
  {journal} {Nano Convergence}\ }\textbf {\bibinfo {volume} {4}},\ \bibinfo
  {pages} {32} (\bibinfo {year} {2017})}\BibitemShut {NoStop}%
\bibitem [{\citenamefont {Bl{\"o}mers}\ \emph {et~al.}(2013)\citenamefont
  {Bl{\"o}mers}, \citenamefont {Rieger}, \citenamefont {Zellekens},
  \citenamefont {Haas}, \citenamefont {Lepsa}, \citenamefont {Hardtdegen},
  \citenamefont {G{\"u}l}, \citenamefont {Demarina}, \citenamefont
  {Gr{\"u}tzmacher}, \citenamefont {L{\"u}th},\ and\ \citenamefont
  {Sch{\"a}pers}}]{Blomers13}%
  \BibitemOpen
  \bibfield  {author} {\bibinfo {author} {\bibfnamefont {C.}~\bibnamefont
  {Bl{\"o}mers}}, \bibinfo {author} {\bibfnamefont {T.}~\bibnamefont {Rieger}},
  \bibinfo {author} {\bibfnamefont {P.}~\bibnamefont {Zellekens}}, \bibinfo
  {author} {\bibfnamefont {F.}~\bibnamefont {Haas}}, \bibinfo {author}
  {\bibfnamefont {M.~I.}\ \bibnamefont {Lepsa}}, \bibinfo {author}
  {\bibfnamefont {H.}~\bibnamefont {Hardtdegen}}, \bibinfo {author}
  {\bibfnamefont {{\"O}.}~\bibnamefont {G{\"u}l}}, \bibinfo {author}
  {\bibfnamefont {N.}~\bibnamefont {Demarina}}, \bibinfo {author}
  {\bibfnamefont {D.}~\bibnamefont {Gr{\"u}tzmacher}}, \bibinfo {author}
  {\bibfnamefont {H.}~\bibnamefont {L{\"u}th}},\ and\ \bibinfo {author}
  {\bibfnamefont {T.}~\bibnamefont {Sch{\"a}pers}},\ }\bibfield  {title}
  {\bibinfo {title} {Realization of nanoscaled tubular conductors by means of
  GaAs/InAs core/shell nanowires},\ }\href@noop {} {\bibfield  {journal}
  {\bibinfo  {journal} {Nanotechnology}\ }\textbf {\bibinfo {volume} {24}},\
  \bibinfo {pages} {035203} (\bibinfo {year} {2013})}\BibitemShut {NoStop}%
\bibitem [{\citenamefont {Rieger}\ \emph {et~al.}(2012)\citenamefont {Rieger},
  \citenamefont {Luysberg}, \citenamefont {Sch{\"a}pers}, \citenamefont
  {Gr{\"u}tzmacher},\ and\ \citenamefont {Lepsa}}]{Rieger12}%
  \BibitemOpen
  \bibfield  {author} {\bibinfo {author} {\bibfnamefont {T.}~\bibnamefont
  {Rieger}}, \bibinfo {author} {\bibfnamefont {M.}~\bibnamefont {Luysberg}},
  \bibinfo {author} {\bibfnamefont {T.}~\bibnamefont {Sch{\"a}pers}}, \bibinfo
  {author} {\bibfnamefont {D.}~\bibnamefont {Gr{\"u}tzmacher}},\ and\ \bibinfo
  {author} {\bibfnamefont {M.~I.}\ \bibnamefont {Lepsa}},\ }\bibfield  {title}
  {\bibinfo {title} {Molecular beam epitaxy growth of GaAs/InAs core-shell
  nanowires and fabrication of InAs nanotubes},\ }\href@noop {} {\bibfield
  {journal} {\bibinfo  {journal} {Nano Letters}\ }\textbf {\bibinfo {volume}
  {12}},\ \bibinfo {pages} {5559} (\bibinfo {year} {2012})}\BibitemShut
  {NoStop}%
\bibitem [{\citenamefont {Haas}\ \emph {et~al.}(2013)\citenamefont {Haas},
  \citenamefont {Sladek}, \citenamefont {Winden}, \citenamefont {von~der Ahe},
  \citenamefont {Weirich}, \citenamefont {Rieger}, \citenamefont {L{\"u}th},
  \citenamefont {Gr{\"u}tzmacher}, \citenamefont {Sch{\"a}pers},\ and\
  \citenamefont {Hardtdegen}}]{Haas13}%
  \BibitemOpen
  \bibfield  {author} {\bibinfo {author} {\bibfnamefont {F.}~\bibnamefont
  {Haas}}, \bibinfo {author} {\bibfnamefont {K.}~\bibnamefont {Sladek}},
  \bibinfo {author} {\bibfnamefont {A.}~\bibnamefont {Winden}}, \bibinfo
  {author} {\bibfnamefont {M.}~\bibnamefont {von~der Ahe}}, \bibinfo {author}
  {\bibfnamefont {T.~E.}\ \bibnamefont {Weirich}}, \bibinfo {author}
  {\bibfnamefont {T.}~\bibnamefont {Rieger}}, \bibinfo {author} {\bibfnamefont
  {H.}~\bibnamefont {L{\"u}th}}, \bibinfo {author} {\bibfnamefont
  {D.}~\bibnamefont {Gr{\"u}tzmacher}}, \bibinfo {author} {\bibfnamefont
  {T.}~\bibnamefont {Sch{\"a}pers}},\ and\ \bibinfo {author} {\bibfnamefont
  {H.}~\bibnamefont {Hardtdegen}},\ }\bibfield  {title} {\bibinfo {title}
  {Nanoimprint and selective-area MOVPE for growth of GaAs/InAs core/shell
  nanowires},\ }\href@noop {} {\bibfield  {journal} {\bibinfo  {journal}
  {Nanotechnology}\ }\textbf {\bibinfo {volume} {24}},\ \bibinfo {pages}
  {085603} (\bibinfo {year} {2013})}\BibitemShut {NoStop}%
\bibitem [{\citenamefont {Fickenscher}\ \emph {et~al.}(2013)\citenamefont
  {Fickenscher}, \citenamefont {Shi}, \citenamefont {Jackson}, \citenamefont
  {Smith}, \citenamefont {Yarrison-Rice}, \citenamefont {Zheng}, \citenamefont
  {Miller}, \citenamefont {Etheridge}, \citenamefont {Wong}, \citenamefont
  {Gao}, \citenamefont {Deshpande}, \citenamefont {Tan},\ and\ \citenamefont
  {Jagadish}}]{Fickenscher13}%
  \BibitemOpen
  \bibfield  {author} {\bibinfo {author} {\bibfnamefont {M.}~\bibnamefont
  {Fickenscher}}, \bibinfo {author} {\bibfnamefont {T.}~\bibnamefont {Shi}},
  \bibinfo {author} {\bibfnamefont {H.~E.}\ \bibnamefont {Jackson}}, \bibinfo
  {author} {\bibfnamefont {L.~M.}\ \bibnamefont {Smith}}, \bibinfo {author}
  {\bibfnamefont {J.~M.}\ \bibnamefont {Yarrison-Rice}}, \bibinfo {author}
  {\bibfnamefont {C.}~\bibnamefont {Zheng}}, \bibinfo {author} {\bibfnamefont
  {P.}~\bibnamefont {Miller}}, \bibinfo {author} {\bibfnamefont
  {J.}~\bibnamefont {Etheridge}}, \bibinfo {author} {\bibfnamefont {B.~M.}\
  \bibnamefont {Wong}}, \bibinfo {author} {\bibfnamefont {Q.}~\bibnamefont
  {Gao}}, \bibinfo {author} {\bibfnamefont {S.}~\bibnamefont {Deshpande}},
  \bibinfo {author} {\bibfnamefont {H.~H.}\ \bibnamefont {Tan}},\ and\ \bibinfo
  {author} {\bibfnamefont {C.}~\bibnamefont {Jagadish}},\ }\bibfield  {title}
  {\bibinfo {title} {Optical, structural, and numerical investigations of
  GaAs/AlGaAs core-multishell nanowire quantum well tubes},\ }\href
  {https://doi.org/10.1021/nl304182j} {\bibfield  {journal} {\bibinfo
  {journal} {Nano Letters}\ }\textbf {\bibinfo {volume} {13}},\ \bibinfo
  {pages} {1016} (\bibinfo {year} {2013})}\BibitemShut {NoStop}%
\bibitem [{\citenamefont {Funk}\ \emph {et~al.}(2013)\citenamefont {Funk},
  \citenamefont {Royo}, \citenamefont {Zardo}, \citenamefont {Rudolph},
  \citenamefont {Morkötter}, \citenamefont {Mayer}, \citenamefont {Becker},
  \citenamefont {Bechtold}, \citenamefont {Matich}, \citenamefont {Döblinger},
  \citenamefont {Bichler}, \citenamefont {Koblmüller}, \citenamefont {Finley},
  \citenamefont {Bertoni}, \citenamefont {Goldoni},\ and\ \citenamefont
  {Abstreiter}}]{Funk13}%
  \BibitemOpen
  \bibfield  {author} {\bibinfo {author} {\bibfnamefont {S.}~\bibnamefont
  {Funk}}, \bibinfo {author} {\bibfnamefont {M.}~\bibnamefont {Royo}}, \bibinfo
  {author} {\bibfnamefont {I.}~\bibnamefont {Zardo}}, \bibinfo {author}
  {\bibfnamefont {D.}~\bibnamefont {Rudolph}}, \bibinfo {author} {\bibfnamefont
  {S.}~\bibnamefont {Morkötter}}, \bibinfo {author} {\bibfnamefont
  {B.}~\bibnamefont {Mayer}}, \bibinfo {author} {\bibfnamefont
  {J.}~\bibnamefont {Becker}}, \bibinfo {author} {\bibfnamefont
  {A.}~\bibnamefont {Bechtold}}, \bibinfo {author} {\bibfnamefont
  {S.}~\bibnamefont {Matich}}, \bibinfo {author} {\bibfnamefont
  {M.}~\bibnamefont {Döblinger}}, \bibinfo {author} {\bibfnamefont
  {M.}~\bibnamefont {Bichler}}, \bibinfo {author} {\bibfnamefont
  {G.}~\bibnamefont {Koblmüller}}, \bibinfo {author} {\bibfnamefont {J.~J.}\
  \bibnamefont {Finley}}, \bibinfo {author} {\bibfnamefont {A.}~\bibnamefont
  {Bertoni}}, \bibinfo {author} {\bibfnamefont {G.}~\bibnamefont {Goldoni}},\
  and\ \bibinfo {author} {\bibfnamefont {G.}~\bibnamefont {Abstreiter}},\
  }\bibfield  {title} {\bibinfo {title} {High mobility one- and two-dimensional
  electron systems in nanowire-based quantum heterostructures},\ }\href@noop {}
  {\bibfield  {journal} {\bibinfo  {journal} {Nano Letters}\ }\textbf {\bibinfo
  {volume} {13}},\ \bibinfo {pages} {6189} (\bibinfo {year}
  {2013})}\BibitemShut {NoStop}%
\bibitem [{\citenamefont {Jadczak}\ \emph {et~al.}(2014)\citenamefont
  {Jadczak}, \citenamefont {Plochocka}, \citenamefont {Mitioglu}, \citenamefont
  {Breslavetz}, \citenamefont {Royo}, \citenamefont {Bertoni}, \citenamefont
  {Goldoni}, \citenamefont {Smolenski}, \citenamefont {Kossacki}, \citenamefont
  {Kretinin}, \citenamefont {Shtrikman},\ and\ \citenamefont
  {Maude}}]{Jadczak14}%
  \BibitemOpen
  \bibfield  {author} {\bibinfo {author} {\bibfnamefont {J.}~\bibnamefont
  {Jadczak}}, \bibinfo {author} {\bibfnamefont {P.}~\bibnamefont {Plochocka}},
  \bibinfo {author} {\bibfnamefont {A.}~\bibnamefont {Mitioglu}}, \bibinfo
  {author} {\bibfnamefont {I.}~\bibnamefont {Breslavetz}}, \bibinfo {author}
  {\bibfnamefont {M.}~\bibnamefont {Royo}}, \bibinfo {author} {\bibfnamefont
  {A.}~\bibnamefont {Bertoni}}, \bibinfo {author} {\bibfnamefont
  {G.}~\bibnamefont {Goldoni}}, \bibinfo {author} {\bibfnamefont
  {T.}~\bibnamefont {Smolenski}}, \bibinfo {author} {\bibfnamefont
  {P.}~\bibnamefont {Kossacki}}, \bibinfo {author} {\bibfnamefont
  {A.}~\bibnamefont {Kretinin}}, \bibinfo {author} {\bibfnamefont
  {H.}~\bibnamefont {Shtrikman}},\ and\ \bibinfo {author} {\bibfnamefont
  {D.~K.}\ \bibnamefont {Maude}},\ }\bibfield  {title} {\bibinfo {title}
  {Unintentional high-density p-type modulation doping of a GaAs/AlAs
  core-multishell nanowire},\ }\href@noop {} {\bibfield  {journal} {\bibinfo
  {journal} {Nano Letters}\ }\textbf {\bibinfo {volume} {14}},\ \bibinfo
  {pages} {2807} (\bibinfo {year} {2014})}\BibitemShut {NoStop}%
\bibitem [{\citenamefont {G\"ul}\ \emph {et~al.}(2014)\citenamefont {G\"ul},
  \citenamefont {Demarina}, \citenamefont {Bl\"omers}, \citenamefont {Rieger},
  \citenamefont {L\"uth}, \citenamefont {Lepsa}, \citenamefont
  {Gr\"utzmacher},\ and\ \citenamefont {Sch\"apers}}]{Gul14}%
  \BibitemOpen
  \bibfield  {author} {\bibinfo {author} {\bibfnamefont {O.}~\bibnamefont
  {G\"ul}}, \bibinfo {author} {\bibfnamefont {N.}~\bibnamefont {Demarina}},
  \bibinfo {author} {\bibfnamefont {C.}~\bibnamefont {Bl\"omers}}, \bibinfo
  {author} {\bibfnamefont {T.}~\bibnamefont {Rieger}}, \bibinfo {author}
  {\bibfnamefont {H.}~\bibnamefont {L\"uth}}, \bibinfo {author} {\bibfnamefont
  {M.~I.}\ \bibnamefont {Lepsa}}, \bibinfo {author} {\bibfnamefont
  {D.}~\bibnamefont {Gr\"utzmacher}},\ and\ \bibinfo {author} {\bibfnamefont
  {T.}~\bibnamefont {Sch\"apers}},\ }\bibfield  {title} {\bibinfo {title} {Flux
  periodic magnetoconductance oscillations in GaAs/InAs core/shell nanowires},\
  }\href@noop {} {\bibfield  {journal} {\bibinfo  {journal} {Phys. Rev. B}\
  }\textbf {\bibinfo {volume} {89}},\ \bibinfo {pages} {045417} (\bibinfo
  {year} {2014})}\BibitemShut {NoStop}%
\bibitem [{\citenamefont {Wei{\ss}}\ \emph {et~al.}(2014)\citenamefont
  {Wei{\ss}}, \citenamefont {Kinzel}, \citenamefont {Sch{\"u}lein},
  \citenamefont {Heigl}, \citenamefont {Rudolph}, \citenamefont
  {Mork{\"o}tter}, \citenamefont {D{\"o}blinger}, \citenamefont {Bichler},
  \citenamefont {Abstreiter}, \citenamefont {Finley}, \citenamefont
  {Koblm{\"u}ller}, \citenamefont {Wixforth},\ and\ \citenamefont
  {Krenner}}]{Weiss14b}%
  \BibitemOpen
  \bibfield  {author} {\bibinfo {author} {\bibfnamefont {M.}~\bibnamefont
  {Wei{\ss}}}, \bibinfo {author} {\bibfnamefont {J.~B.}\ \bibnamefont
  {Kinzel}}, \bibinfo {author} {\bibfnamefont {F.~J.~R.}\ \bibnamefont
  {Sch{\"u}lein}}, \bibinfo {author} {\bibfnamefont {M.}~\bibnamefont {Heigl}},
  \bibinfo {author} {\bibfnamefont {D.}~\bibnamefont {Rudolph}}, \bibinfo
  {author} {\bibfnamefont {S.}~\bibnamefont {Mork{\"o}tter}}, \bibinfo {author}
  {\bibfnamefont {M.}~\bibnamefont {D{\"o}blinger}}, \bibinfo {author}
  {\bibfnamefont {M.}~\bibnamefont {Bichler}}, \bibinfo {author} {\bibfnamefont
  {G.}~\bibnamefont {Abstreiter}}, \bibinfo {author} {\bibfnamefont {J.~J.}\
  \bibnamefont {Finley}}, \bibinfo {author} {\bibfnamefont {G.}~\bibnamefont
  {Koblm{\"u}ller}}, \bibinfo {author} {\bibfnamefont {A.}~\bibnamefont
  {Wixforth}},\ and\ \bibinfo {author} {\bibfnamefont {H.~J.}\ \bibnamefont
  {Krenner}},\ }\bibfield  {title} {\bibinfo {title} {Dynamic acoustic control
  of individual optically active quantum dot-like emission centers in
  heterostructure nanowires},\ }\href@noop {} {\bibfield  {journal} {\bibinfo
  {journal} {Nano Letters}\ }\textbf {\bibinfo {volume} {14}},\ \bibinfo
  {pages} {2256} (\bibinfo {year} {2014})}\BibitemShut {NoStop}%
\bibitem [{\citenamefont {Erhard}\ \emph {et~al.}(2015)\citenamefont {Erhard},
  \citenamefont {Zenger}, \citenamefont {Morkötter}, \citenamefont {Rudolph},
  \citenamefont {Weiss}, \citenamefont {Krenner}, \citenamefont {Karl},
  \citenamefont {Abstreiter}, \citenamefont {Finley}, \citenamefont
  {Koblm{\"u}ller},\ and\ \citenamefont {Holleitner}}]{Erhard15}%
  \BibitemOpen
  \bibfield  {author} {\bibinfo {author} {\bibfnamefont {N.}~\bibnamefont
  {Erhard}}, \bibinfo {author} {\bibfnamefont {S.}~\bibnamefont {Zenger}},
  \bibinfo {author} {\bibfnamefont {S.}~\bibnamefont {Morkötter}}, \bibinfo
  {author} {\bibfnamefont {D.}~\bibnamefont {Rudolph}}, \bibinfo {author}
  {\bibfnamefont {M.}~\bibnamefont {Weiss}}, \bibinfo {author} {\bibfnamefont
  {H.~J.}\ \bibnamefont {Krenner}}, \bibinfo {author} {\bibfnamefont
  {H.}~\bibnamefont {Karl}}, \bibinfo {author} {\bibfnamefont {G.}~\bibnamefont
  {Abstreiter}}, \bibinfo {author} {\bibfnamefont {J.~J.}\ \bibnamefont
  {Finley}}, \bibinfo {author} {\bibfnamefont {G.}~\bibnamefont
  {Koblm{\"u}ller}},\ and\ \bibinfo {author} {\bibfnamefont {A.~W.}\
  \bibnamefont {Holleitner}},\ }\bibfield  {title} {\bibinfo {title} {Ultrafast
  photodetection in the quantum wells of single AlGaAs/GaAs-based nanowires},\
  }\href@noop {} {\bibfield  {journal} {\bibinfo  {journal} {Nano Letters}\
  }\textbf {\bibinfo {volume} {15}},\ \bibinfo {pages} {6869} (\bibinfo {year}
  {2015})}\BibitemShut {NoStop}%
\bibitem [{\citenamefont {Shi}\ \emph {et~al.}(2015)\citenamefont {Shi},
  \citenamefont {Jackson}, \citenamefont {Smith}, \citenamefont {Jiang},
  \citenamefont {Gao}, \citenamefont {Tan}, \citenamefont {Jagadish},
  \citenamefont {Zheng},\ and\ \citenamefont {Etheridge}}]{Shi15}%
  \BibitemOpen
  \bibfield  {author} {\bibinfo {author} {\bibfnamefont {T.}~\bibnamefont
  {Shi}}, \bibinfo {author} {\bibfnamefont {H.~E.}\ \bibnamefont {Jackson}},
  \bibinfo {author} {\bibfnamefont {L.~M.}\ \bibnamefont {Smith}}, \bibinfo
  {author} {\bibfnamefont {N.}~\bibnamefont {Jiang}}, \bibinfo {author}
  {\bibfnamefont {Q.}~\bibnamefont {Gao}}, \bibinfo {author} {\bibfnamefont
  {H.~H.}\ \bibnamefont {Tan}}, \bibinfo {author} {\bibfnamefont
  {C.}~\bibnamefont {Jagadish}}, \bibinfo {author} {\bibfnamefont
  {C.}~\bibnamefont {Zheng}},\ and\ \bibinfo {author} {\bibfnamefont
  {J.}~\bibnamefont {Etheridge}},\ }\bibfield  {title} {\bibinfo {title}
  {Emergence of localized states in narrow GaAs/AlGaAs nanowire quantum well
  tubes},\ }\href {https://doi.org/10.1021/nl5046878} {\bibfield  {journal}
  {\bibinfo  {journal} {Nano Letters}\ }\textbf {\bibinfo {volume} {15}},\
  \bibinfo {pages} {1876} (\bibinfo {year} {2015})}\BibitemShut {NoStop}%
\bibitem [{\citenamefont {Sonner}\ \emph {et~al.}(2019)\citenamefont {Sonner},
  \citenamefont {Sitek}, \citenamefont {Janker}, \citenamefont {Rudolph},
  \citenamefont {Ruhstorfer}, \citenamefont {D{\"o}blinger}, \citenamefont
  {Manolescu}, \citenamefont {Abstreiter}, \citenamefont {Finley},
  \citenamefont {Wixforth}, \citenamefont {Koblmüller},\ and\ \citenamefont
  {Krenner}}]{Sonner19}%
  \BibitemOpen
  \bibfield  {author} {\bibinfo {author} {\bibfnamefont {M.~M.}\ \bibnamefont
  {Sonner}}, \bibinfo {author} {\bibfnamefont {A.}~\bibnamefont {Sitek}},
  \bibinfo {author} {\bibfnamefont {L.}~\bibnamefont {Janker}}, \bibinfo
  {author} {\bibfnamefont {D.}~\bibnamefont {Rudolph}}, \bibinfo {author}
  {\bibfnamefont {D.}~\bibnamefont {Ruhstorfer}}, \bibinfo {author}
  {\bibfnamefont {M.}~\bibnamefont {D{\"o}blinger}}, \bibinfo {author}
  {\bibfnamefont {A.}~\bibnamefont {Manolescu}}, \bibinfo {author}
  {\bibfnamefont {G.}~\bibnamefont {Abstreiter}}, \bibinfo {author}
  {\bibfnamefont {J.~J.}\ \bibnamefont {Finley}}, \bibinfo {author}
  {\bibfnamefont {A.}~\bibnamefont {Wixforth}}, \bibinfo {author}
  {\bibfnamefont {G.}~\bibnamefont {Koblmüller}},\ and\ \bibinfo {author}
  {\bibfnamefont {H.~J.}\ \bibnamefont {Krenner}},\ }\bibfield  {title}
  {\bibinfo {title} {Breakdown of corner states and carrier localization by
  monolayer fluctuations in radial nanowire quantum wells},\ } {\bibfield  {journal}
  {\bibinfo  {journal} {Nano Letters}\ }\textbf {\bibinfo {volume} {19}},\
  \bibinfo {pages} {3336} (\bibinfo {year} {2019})} \BibitemShut {NoStop}%
\bibitem [{\citenamefont {Qian}\ \emph {et~al.}(2004)\citenamefont {Qian},
  \citenamefont {Li}, \citenamefont {Grade{\v{c}}ak}, \citenamefont {Wang},
  \citenamefont {Barrelet},\ and\ \citenamefont {Lieber}}]{Qian04}%
  \BibitemOpen
  \bibfield  {author} {\bibinfo {author} {\bibfnamefont {F.}~\bibnamefont
  {Qian}}, \bibinfo {author} {\bibfnamefont {Y.}~\bibnamefont {Li}}, \bibinfo
  {author} {\bibfnamefont {S.}~\bibnamefont {Grade{\v{c}}ak}}, \bibinfo
  {author} {\bibfnamefont {D.}~\bibnamefont {Wang}}, \bibinfo {author}
  {\bibfnamefont {C.~J.}\ \bibnamefont {Barrelet}},\ and\ \bibinfo {author}
  {\bibfnamefont {C.~M.}\ \bibnamefont {Lieber}},\ }\bibfield  {title}
  {\bibinfo {title} {Gallium nitride-based nanowire radial heterostructures for
  nanophotonics},\ }\href@noop {} {\bibfield  {journal} {\bibinfo  {journal}
  {Nano Letters}\ }\textbf {\bibinfo {volume} {4}},\ \bibinfo {pages} {1975}
  (\bibinfo {year} {2004})}\BibitemShut {NoStop}%
\bibitem [{\citenamefont {Qian}\ \emph {et~al.}(2005)\citenamefont {Qian},
  \citenamefont {Grade{\v{c}}ak}, \citenamefont {Li}, \citenamefont {Wen},\
  and\ \citenamefont {Lieber}}]{Qian05}%
  \BibitemOpen
  \bibfield  {author} {\bibinfo {author} {\bibfnamefont {F.}~\bibnamefont
  {Qian}}, \bibinfo {author} {\bibfnamefont {S.}~\bibnamefont
  {Grade{\v{c}}ak}}, \bibinfo {author} {\bibfnamefont {Y.}~\bibnamefont {Li}},
  \bibinfo {author} {\bibfnamefont {C.-Y.}\ \bibnamefont {Wen}},\ and\ \bibinfo
  {author} {\bibfnamefont {C.~M.}\ \bibnamefont {Lieber}},\ }\bibfield  {title}
  {\bibinfo {title} {Core/multishell nanowire heterostructures as multicolor,
  high-efficiency light-emitting diodes},\ }\href@noop {} {\bibfield  {journal}
  {\bibinfo  {journal} {Nano Letters}\ }\textbf {\bibinfo {volume} {5}},\
  \bibinfo {pages} {2287} (\bibinfo {year} {2005})}\BibitemShut {NoStop}%
\bibitem [{\citenamefont {Baird}\ \emph {et~al.}(2009)\citenamefont {Baird},
  \citenamefont {Ang}, \citenamefont {Low}, \citenamefont {Haegel},
  \citenamefont {Talin}, \citenamefont {Li},\ and\ \citenamefont
  {Wang}}]{Baird09}%
  \BibitemOpen
  \bibfield  {author} {\bibinfo {author} {\bibfnamefont {L.}~\bibnamefont
  {Baird}}, \bibinfo {author} {\bibfnamefont {G.}~\bibnamefont {Ang}}, \bibinfo
  {author} {\bibfnamefont {C.}~\bibnamefont {Low}}, \bibinfo {author}
  {\bibfnamefont {N.}~\bibnamefont {Haegel}}, \bibinfo {author} {\bibfnamefont
  {A.}~\bibnamefont {Talin}}, \bibinfo {author} {\bibfnamefont
  {Q.}~\bibnamefont {Li}},\ and\ \bibinfo {author} {\bibfnamefont
  {G.}~\bibnamefont {Wang}},\ }\bibfield  {title} {\bibinfo {title} {Imaging
  minority carrier diffusion in GaN nanowires using near field optical
  microscopy},\ }\href@noop {} {\bibfield  {journal} {\bibinfo  {journal}
  {Physica B: Condensed Matter}\ }\textbf {\bibinfo {volume} {404}},\ \bibinfo
  {pages} {4933 } (\bibinfo {year} {2009})}\BibitemShut {NoStop}%
\bibitem [{\citenamefont {Heurlin}\ \emph {et~al.}(2015)\citenamefont
  {Heurlin}, \citenamefont {Stankevi{\v{c}}}, \citenamefont
  {Mickevi{\v{c}}ius}, \citenamefont {Yngman}, \citenamefont {Lindgren},
  \citenamefont {Mikkelsen}, \citenamefont {Feidenhans’l}, \citenamefont
  {Borgst{\"o}m},\ and\ \citenamefont {Samuelson}}]{Heurlin15}%
  \BibitemOpen
  \bibfield  {author} {\bibinfo {author} {\bibfnamefont {M.}~\bibnamefont
  {Heurlin}}, \bibinfo {author} {\bibfnamefont {T.}~\bibnamefont
  {Stankevi{\v{c}}}}, \bibinfo {author} {\bibfnamefont {S.}~\bibnamefont
  {Mickevi{\v{c}}ius}}, \bibinfo {author} {\bibfnamefont {S.}~\bibnamefont
  {Yngman}}, \bibinfo {author} {\bibfnamefont {D.}~\bibnamefont {Lindgren}},
  \bibinfo {author} {\bibfnamefont {A.}~\bibnamefont {Mikkelsen}}, \bibinfo
  {author} {\bibfnamefont {R.}~\bibnamefont {Feidenhans’l}}, \bibinfo
  {author} {\bibfnamefont {M.~T.}\ \bibnamefont {Borgst{\"o}m}},\ and\ \bibinfo
  {author} {\bibfnamefont {L.}~\bibnamefont {Samuelson}},\ }\bibfield  {title}
  {\bibinfo {title} {Structural properties of wurtzite InP-InGaAs nanowire
  core-shell heterostructures},\ }\href@noop {} {\bibfield  {journal}
  {\bibinfo  {journal} {Nano Letters}\ }\textbf {\bibinfo {volume} {15}},\
  \bibinfo {pages} {2462} (\bibinfo {year} {2015})}\BibitemShut {NoStop}%
\bibitem [{\citenamefont {Dong}\ \emph {et~al.}(2009)\citenamefont {Dong},
  \citenamefont {Tian}, \citenamefont {Kempa},\ and\ \citenamefont
  {Lieber}}]{Dong09}%
  \BibitemOpen
  \bibfield  {author} {\bibinfo {author} {\bibfnamefont {Y.}~\bibnamefont
  {Dong}}, \bibinfo {author} {\bibfnamefont {B.}~\bibnamefont {Tian}}, \bibinfo
  {author} {\bibfnamefont {T.~J.}\ \bibnamefont {Kempa}},\ and\ \bibinfo
  {author} {\bibfnamefont {C.~M.}\ \bibnamefont {Lieber}},\ }\bibfield  {title}
  {\bibinfo {title} {Coaxial group III-nitride nanowire photovoltaics},\
  }\href@noop {} {\bibfield  {journal} {\bibinfo  {journal} {Nano Letters}\
  }\textbf {\bibinfo {volume} {9}},\ \bibinfo {pages} {2183} (\bibinfo {year}
  {2009})}\BibitemShut {NoStop}%
\bibitem [{\citenamefont {G\"uniat}\ \emph {et~al.}(2019)\citenamefont
  {G\"uniat}, \citenamefont {Martí-Sánchez}, \citenamefont {Garcia},
  \citenamefont {Boscardin}, \citenamefont {Vindice}, \citenamefont {Tappy},
  \citenamefont {Friedl}, \citenamefont {Kim}, \citenamefont {Zamani},
  \citenamefont {Francaviglia}, \citenamefont {Balgarkashi}, \citenamefont
  {Leran}, \citenamefont {Arbiol},\ and\ \citenamefont {Fontcuberta~i
  Morral}}]{Guniat19}%
  \BibitemOpen
  \bibfield  {author} {\bibinfo {author} {\bibfnamefont {L.}~\bibnamefont
  {G\"uniat}}, \bibinfo {author} {\bibfnamefont {S.}~\bibnamefont
  {Martí-Sánchez}}, \bibinfo {author} {\bibfnamefont {O.}~\bibnamefont
  {Garcia}}, \bibinfo {author} {\bibfnamefont {M.}~\bibnamefont {Boscardin}},
  \bibinfo {author} {\bibfnamefont {D.}~\bibnamefont {Vindice}}, \bibinfo
  {author} {\bibfnamefont {N.}~\bibnamefont {Tappy}}, \bibinfo {author}
  {\bibfnamefont {M.}~\bibnamefont {Friedl}}, \bibinfo {author} {\bibfnamefont
  {W.}~\bibnamefont {Kim}}, \bibinfo {author} {\bibfnamefont {M.}~\bibnamefont
  {Zamani}}, \bibinfo {author} {\bibfnamefont {L.}~\bibnamefont
  {Francaviglia}}, \bibinfo {author} {\bibfnamefont {A.}~\bibnamefont
  {Balgarkashi}}, \bibinfo {author} {\bibfnamefont {J.-B.}\ \bibnamefont
  {Leran}}, \bibinfo {author} {\bibfnamefont {J.}~\bibnamefont {Arbiol}},\ and\
  \bibinfo {author} {\bibfnamefont {A.}~\bibnamefont {Fontcuberta~i Morral}},\
  }\bibfield  {title} {\bibinfo {title} {III-V integration on Si(100):
  Vertical nanospades},\ }\href {https://doi.org/10.1021/acsnano.9b01546}
  {\bibfield  {journal} {\bibinfo  {journal} {ACS Nano}\ }\textbf {\bibinfo
  {volume} {13}},\ \bibinfo {pages} {5833} (\bibinfo {year}
  {2019})}\BibitemShut {NoStop}%
\bibitem [{\citenamefont {Fonseka}\ \emph {et~al.}(2019)\citenamefont
  {Fonseka}, \citenamefont {Caroff}, \citenamefont {Guo}, \citenamefont
  {Sanchez}, \citenamefont {Tan},\ and\ \citenamefont {Jagadish}}]{Fonseka19}%
  \BibitemOpen
  \bibfield  {author} {\bibinfo {author} {\bibfnamefont {H.~A.}\ \bibnamefont
  {Fonseka}}, \bibinfo {author} {\bibfnamefont {P.}~\bibnamefont {Caroff}},
  \bibinfo {author} {\bibfnamefont {Y.}~\bibnamefont {Guo}}, \bibinfo {author}
  {\bibfnamefont {A.~M.}\ \bibnamefont {Sanchez}}, \bibinfo {author}
  {\bibfnamefont {H.~H.}\ \bibnamefont {Tan}},\ and\ \bibinfo {author}
  {\bibfnamefont {C.}~\bibnamefont {Jagadish}},\ }\bibfield  {title} {\bibinfo
  {title} {Engineering the side facets of vertical [100] oriented InP nanowires
  for novel radial heterostructures},\ }\href
  {https://doi.org/10.1186/s11671-019-3177-6} {\bibfield  {journal} {\bibinfo
  {journal} {Nanoscale Research Letters}\ }\textbf {\bibinfo {volume} {14}},\
  \bibinfo {pages} {399} (\bibinfo {year} {2019})}\BibitemShut {NoStop}%
\bibitem [{\citenamefont {Rieger}\ \emph {et~al.}(2015)\citenamefont {Rieger},
  \citenamefont {Grutzmacher},\ and\ \citenamefont {Lepsa}}]{Rieger15}%
  \BibitemOpen
  \bibfield  {author} {\bibinfo {author} {\bibfnamefont {T.}~\bibnamefont
  {Rieger}}, \bibinfo {author} {\bibfnamefont {D.}~\bibnamefont
  {Grutzmacher}},\ and\ \bibinfo {author} {\bibfnamefont {M.~I.}\ \bibnamefont
  {Lepsa}},\ }\bibfield  {title} {\bibinfo {title} {Misfit dislocation free
  InAs/GaSb core-shell nanowires grown by molecular beam epitaxy},\ }\href@noop
  {} {\bibfield  {journal} {\bibinfo  {journal} {Nanoscale}\ }\textbf {\bibinfo
  {volume} {7}},\ \bibinfo {pages} {356} (\bibinfo {year} {2015})}\BibitemShut
  {NoStop}%
\bibitem [{\citenamefont {Yuan}\ \emph {et~al.}(2015)\citenamefont {Yuan},
  \citenamefont {Caroff}, \citenamefont {Wang}, \citenamefont {Guo},
  \citenamefont {Wang}, \citenamefont {Jackson}, \citenamefont {Smith},
  \citenamefont {Tan},\ and\ \citenamefont {Jagadish}}]{Yuan15}%
  \BibitemOpen
  \bibfield  {author} {\bibinfo {author} {\bibfnamefont {X.}~\bibnamefont
  {Yuan}}, \bibinfo {author} {\bibfnamefont {P.}~\bibnamefont {Caroff}},
  \bibinfo {author} {\bibfnamefont {F.}~\bibnamefont {Wang}}, \bibinfo {author}
  {\bibfnamefont {Y.}~\bibnamefont {Guo}}, \bibinfo {author} {\bibfnamefont
  {Y.}~\bibnamefont {Wang}}, \bibinfo {author} {\bibfnamefont {H.~E.}\
  \bibnamefont {Jackson}}, \bibinfo {author} {\bibfnamefont {L.~M.}\
  \bibnamefont {Smith}}, \bibinfo {author} {\bibfnamefont {H.~H.}\ \bibnamefont
  {Tan}},\ and\ \bibinfo {author} {\bibfnamefont {C.}~\bibnamefont
  {Jagadish}},\ }\bibfield  {title} {\bibinfo {title} {Antimony induced \{112\}A
  faceted triangular GaAs$_{1-x}$Sb$_{x}$/InP core/shell nanowires and their
  enhanced optical quality},\ }\href@noop {} {\bibfield  {journal} {\bibinfo
  {journal} {Adv. Funct. Mater.}\ }\textbf {\bibinfo {volume} {25}},\ \bibinfo
  {pages} {5300} (\bibinfo {year} {2015})}\BibitemShut {NoStop}%
\bibitem [{\citenamefont {G\"{o}ransson}\ \emph {et~al.}(2019)\citenamefont
  {G\"{o}ransson}, \citenamefont {Heurlin}, \citenamefont {Dalelkhan},
  \citenamefont {Abay}, \citenamefont {Messing}, \citenamefont {Maisi},
  \citenamefont {Borgström},\ and\ \citenamefont {Xu}}]{Goransson19}%
  \BibitemOpen
  \bibfield  {author} {\bibinfo {author} {\bibfnamefont {D.~J.~O.}\
  \bibnamefont {G\"{o}ransson}}, \bibinfo {author} {\bibfnamefont
  {M.}~\bibnamefont {Heurlin}}, \bibinfo {author} {\bibfnamefont
  {B.}~\bibnamefont {Dalelkhan}}, \bibinfo {author} {\bibfnamefont
  {S.}~\bibnamefont {Abay}}, \bibinfo {author} {\bibfnamefont {M.~E.}\
  \bibnamefont {Messing}}, \bibinfo {author} {\bibfnamefont {V.~F.}\
  \bibnamefont {Maisi}}, \bibinfo {author} {\bibfnamefont {M.~T.}\ \bibnamefont
  {Borgström}},\ and\ \bibinfo {author} {\bibfnamefont {H.~Q.}\ \bibnamefont
  {Xu}},\ }\bibfield  {title} {\bibinfo {title} {Coulomb blockade from the
  shell of an InP-InAs core-shell nanowire with a triangular cross section},\
  }\href {https://doi.org/10.1063/1.5084222} {\bibfield  {journal} {\bibinfo
  {journal} {Applied Physics Letters}\ }\textbf {\bibinfo {volume} {114}},\
  \bibinfo {pages} {053108} (\bibinfo {year} {2019})}\BibitemShut {NoStop}%
\bibitem [{\citenamefont {Bertoni}\ \emph {et~al.}(2011)\citenamefont
  {Bertoni}, \citenamefont {Royo}, \citenamefont {Mahawish},\ and\
  \citenamefont {Goldoni}}]{Bertoni11}%
  \BibitemOpen
  \bibfield  {author} {\bibinfo {author} {\bibfnamefont {A.}~\bibnamefont
  {Bertoni}}, \bibinfo {author} {\bibfnamefont {M.}~\bibnamefont {Royo}},
  \bibinfo {author} {\bibfnamefont {F.}~\bibnamefont {Mahawish}},\ and\
  \bibinfo {author} {\bibfnamefont {G.}~\bibnamefont {Goldoni}},\ }\bibfield
  {title} {\bibinfo {title} {Electron and hole gas in modulation-doped
  GaAs/Al$_{\mathrm{1-x}}$Ga$_\mathrm{x}$As radial heterojunctions},\ }\href
  {https://doi.org/10.1103/PhysRevB.84.205323} {\bibfield  {journal} {\bibinfo
  {journal} {Phys. Rev. B}\ }\textbf {\bibinfo {volume} {84}},\ \bibinfo
  {pages} {205323} (\bibinfo {year} {2011})}\BibitemShut {NoStop}%
\bibitem [{\citenamefont {Royo}\ \emph {et~al.}(2013)\citenamefont {Royo},
  \citenamefont {Bertoni},\ and\ \citenamefont {Goldoni}}]{Royo13}%
  \BibitemOpen
  \bibfield  {author} {\bibinfo {author} {\bibfnamefont {M.}~\bibnamefont
  {Royo}}, \bibinfo {author} {\bibfnamefont {A.}~\bibnamefont {Bertoni}},\ and\
  \bibinfo {author} {\bibfnamefont {G.}~\bibnamefont {Goldoni}},\ }\bibfield
  {title} {\bibinfo {title} {Landau levels, edge states, and magnetoconductance
  in GaAs/AlGaAs core-shell nanowires},\ }\href
  {https://doi.org/10.1103/PhysRevB.87.115316} {\bibfield  {journal} {\bibinfo
  {journal} {Phys. Rev. B}\ }\textbf {\bibinfo {volume} {87}},\ \bibinfo
  {pages} {115316} (\bibinfo {year} {2013})}\BibitemShut {NoStop}%
\bibitem [{\citenamefont {Wong}\ \emph {et~al.}(2011)\citenamefont {Wong},
  \citenamefont {Léonard}, \citenamefont {Li},\ and\ \citenamefont
  {Wang}}]{Wong11}%
  \BibitemOpen
  \bibfield  {author} {\bibinfo {author} {\bibfnamefont {B.~M.}\ \bibnamefont
  {Wong}}, \bibinfo {author} {\bibfnamefont {F.}~\bibnamefont {Léonard}},
  \bibinfo {author} {\bibfnamefont {Q.}~\bibnamefont {Li}},\ and\ \bibinfo
  {author} {\bibfnamefont {G.~T.}\ \bibnamefont {Wang}},\ }\bibfield  {title}
  {\bibinfo {title} {Nanoscale effects on heterojunction electron gases in
  GaN/AlGaN core/shell nanowires},\ }\href@noop {} {\bibfield  {journal}
  {\bibinfo  {journal} {Nano Letters}\ }\textbf {\bibinfo {volume} {11}},\
  \bibinfo {pages} {3074} (\bibinfo {year} {2011})}\BibitemShut {NoStop}%
\bibitem [{\citenamefont {Woods}\ \emph {et~al.}(2019)\citenamefont {Woods},
  \citenamefont {Das~Sarma},\ and\ \citenamefont {Stanescu}}]{Woods19}%
  \BibitemOpen
  \bibfield  {author} {\bibinfo {author} {\bibfnamefont {B.~D.}\ \bibnamefont
  {Woods}}, \bibinfo {author} {\bibfnamefont {S.}~\bibnamefont {Das~Sarma}},\
  and\ \bibinfo {author} {\bibfnamefont {T.~D.}\ \bibnamefont {Stanescu}},\
  }\bibfield  {title} {\bibinfo {title} {Electronic structure of full-shell
  InAs/Al hybrid semiconductor-superconductor nanowires: Spin-orbit coupling
  and topological phase space},\ }\href
  {https://doi.org/10.1103/PhysRevB.99.161118} {\bibfield  {journal} {\bibinfo
  {journal} {Phys. Rev. B}\ }\textbf {\bibinfo {volume} {99}},\ \bibinfo
  {pages} {161118} (\bibinfo {year} {2019})}\BibitemShut {NoStop}%
\bibitem [{\citenamefont {Mohan}\ \emph {et~al.}(2006)\citenamefont {Mohan},
  \citenamefont {Motohisa},\ and\ \citenamefont {Fukui}}]{Mohan06}%
  \BibitemOpen
  \bibfield  {author} {\bibinfo {author} {\bibfnamefont {P.}~\bibnamefont
  {Mohan}}, \bibinfo {author} {\bibfnamefont {J.}~\bibnamefont {Motohisa}},\
  and\ \bibinfo {author} {\bibfnamefont {T.}~\bibnamefont {Fukui}},\ }\bibfield
   {title} {\bibinfo {title} {Fabrication of InP/InAs/InP core-multishell
  heterostructure nanowires by selective area metalorganic vapor phase
  epitaxy},\ }\href {https://doi.org/10.1063/1.2189203} {\bibfield  {journal}
  {\bibinfo  {journal} {Applied Physics Letters}\ }\textbf {\bibinfo {volume}
  {88}},\ \bibinfo {pages} {133105} (\bibinfo {year} {2006})}\BibitemShut
  {NoStop}%
\bibitem [{\citenamefont {Kriegner}\ \emph {et~al.}(2011)\citenamefont
  {Kriegner}, \citenamefont {Wintersberger}, \citenamefont {Kawaguchi},
  \citenamefont {Wallentin}, \citenamefont {Borgström},\ and\ \citenamefont
  {Stangl}}]{Kriegner11}%
  \BibitemOpen
  \bibfield  {author} {\bibinfo {author} {\bibfnamefont {D.}~\bibnamefont
  {Kriegner}}, \bibinfo {author} {\bibfnamefont {E.}~\bibnamefont
  {Wintersberger}}, \bibinfo {author} {\bibfnamefont {K.}~\bibnamefont
  {Kawaguchi}}, \bibinfo {author} {\bibfnamefont {J.}~\bibnamefont
  {Wallentin}}, \bibinfo {author} {\bibfnamefont {M.~T.}\ \bibnamefont
  {Borgström}},\ and\ \bibinfo {author} {\bibfnamefont {J.}~\bibnamefont
  {Stangl}},\ }\bibfield  {title} {\bibinfo {title} {Unit cell parameters of
  wurtzite {InP} nanowires determined by x-ray diffraction},\ }\href
  {https://doi.org/10.1088/0957-4484/22/42/425704} {\bibfield  {journal}
  {\bibinfo  {journal} {Nanotechnology}\ }\textbf {\bibinfo {volume} {22}},\
  \bibinfo {pages} {425704} (\bibinfo {year} {2011})}\BibitemShut {NoStop}%
\bibitem [{\citenamefont {Luo}\ \emph {et~al.}(2011)\citenamefont {Luo},
  \citenamefont {Zhang},\ and\ \citenamefont {Zunger}}]{Luo11}%
  \BibitemOpen
  \bibfield  {author} {\bibinfo {author} {\bibfnamefont {J.-W.}\ \bibnamefont
  {Luo}}, \bibinfo {author} {\bibfnamefont {L.}~\bibnamefont {Zhang}},\ and\
  \bibinfo {author} {\bibfnamefont {A.}~\bibnamefont {Zunger}},\ }\bibfield
  {title} {\bibinfo {title} {Absence of intrinsic spin splitting in
  one-dimensional quantum wires of tetrahedral semiconductors},\ }\href
  {https://doi.org/10.1103/PhysRevB.84.121303} {\bibfield  {journal} {\bibinfo
  {journal} {Phys. Rev. B}\ }\textbf {\bibinfo {volume} {84}},\ \bibinfo
  {pages} {121303} (\bibinfo {year} {2011})}\BibitemShut {NoStop}%
\bibitem [{\citenamefont {Peña}\ \emph {et~al.}(2024)\citenamefont {Peña},
  \citenamefont {Koepke}, \citenamefont {Dycus}, \citenamefont {Mounce},
  \citenamefont {Baczewski}, \citenamefont {Jacobson},\ and\ \citenamefont
  {Bussmann}}]{Fabian24}%
  \BibitemOpen
  \bibfield  {author} {\bibinfo {author} {\bibfnamefont {L.~F.}\ \bibnamefont
  {Peña}}, \bibinfo {author} {\bibfnamefont {J.~C.}\ \bibnamefont {Koepke}},
  \bibinfo {author} {\bibfnamefont {J.~H.}\ \bibnamefont {Dycus}}, \bibinfo
  {author} {\bibfnamefont {A.}~\bibnamefont {Mounce}}, \bibinfo {author}
  {\bibfnamefont {A.~D.}\ \bibnamefont {Baczewski}}, \bibinfo {author}
  {\bibfnamefont {N.~T.}\ \bibnamefont {Jacobson}},\ and\ \bibinfo {author}
  {\bibfnamefont {E.}~\bibnamefont {Bussmann}},\ }\bibfield  {title} {\bibinfo
  {title} {Modeling Si/SiGe quantum dot variability induced by interface
  disorder reconstructed from multiperspective microscopy},\ }\href@noop {}
  {\bibfield  {journal} {\bibinfo  {journal} {npj Quantum Information}\
  }\textbf {\bibinfo {volume} {10}} (\bibinfo {year} {2024})}\BibitemShut
  {NoStop}%
\bibitem [{\citenamefont {Vurgaftman}\ \emph {et~al.}(2001)\citenamefont
  {Vurgaftman}, \citenamefont {Meyer},\ and\ \citenamefont
  {Ram-Mohan}}]{Vurgaftman01}%
  \BibitemOpen
  \bibfield  {author} {\bibinfo {author} {\bibfnamefont {I.}~\bibnamefont
  {Vurgaftman}}, \bibinfo {author} {\bibfnamefont {J.~R.}\ \bibnamefont
  {Meyer}},\ and\ \bibinfo {author} {\bibfnamefont {L.~R.}\ \bibnamefont
  {Ram-Mohan}},\ }\bibfield  {title} {\bibinfo {title} {Band parameters for
  III-V compound semiconductors and their alloys},\ }\href
  {https://doi.org/10.1063/1.1368156} {\bibfield  {journal} {\bibinfo
  {journal} {Journal of Applied Physics}\ }\textbf {\bibinfo {volume} {89}},\
  \bibinfo {pages} {5815} (\bibinfo {year} {2001})} \BibitemShut {NoStop}%
\bibitem [{\citenamefont {Sitek}\ \emph {et~al.}(2015)\citenamefont {Sitek},
  \citenamefont {Serra}, \citenamefont {Gudmundsson},\ and\ \citenamefont
  {Manolescu}}]{Sitek15}%
  \BibitemOpen
  \bibfield  {author} {\bibinfo {author} {\bibfnamefont {A.}~\bibnamefont
  {Sitek}}, \bibinfo {author} {\bibfnamefont {L.}~\bibnamefont {Serra}},
  \bibinfo {author} {\bibfnamefont {V.}~\bibnamefont {Gudmundsson}},\ and\
  \bibinfo {author} {\bibfnamefont {A.}~\bibnamefont {Manolescu}},\ }\bibfield
  {title} {\bibinfo {title} {Electron localization and optical absorption of
  polygonal quantum rings},\ }\href@noop {} {\bibfield  {journal} {\bibinfo
  {journal} {Phys. Rev. B}\ }\textbf {\bibinfo {volume} {91}},\ \bibinfo
  {pages} {235429} (\bibinfo {year} {2015})}\BibitemShut {NoStop}%
\bibitem [{\citenamefont {Fabian}\ \emph {et~al.}(2007)\citenamefont {Fabian},
  \citenamefont {Matos-Abiague}, \citenamefont {Ertler}, \citenamefont
  {Stano},\ and\ \citenamefont {Žutić}}]{Fabian07}%
  \BibitemOpen
  \bibfield  {author} {\bibinfo {author} {\bibfnamefont {J.}~\bibnamefont
  {Fabian}}, \bibinfo {author} {\bibfnamefont {A.}~\bibnamefont
  {Matos-Abiague}}, \bibinfo {author} {\bibfnamefont {C.}~\bibnamefont
  {Ertler}}, \bibinfo {author} {\bibfnamefont {P.}~\bibnamefont {Stano}},\ and\
  \bibinfo {author} {\bibfnamefont {I.}~\bibnamefont {Žutić}},\ }\bibfield
  {title} {\bibinfo {title} {Semiconductor spintronics},\ }\href@noop {}
  {\bibfield  {journal} {\bibinfo  {journal} {Acta Phys. Slovaca}\ }\textbf
  {\bibinfo {volume} {57}},\ \bibinfo {pages} {565} (\bibinfo {year}
  {2007})}\BibitemShut {NoStop}%
\bibitem [{\citenamefont {Winkler}(2003)}]{Winkler}%
  \BibitemOpen
  \bibfield  {author} {\bibinfo {author} {\bibfnamefont {R.}~\bibnamefont
  {Winkler}},\ }\href {https://doi.org/10.1007/b13586} {\emph {\bibinfo {title}
  {Spin-Orbit Effects in Two-Dimensional Electron and Hole Systems},}}\
  \bibinfo  {publisher} {Springer Tracts in Modern Physics (Springer, Berlin},\ \bibinfo {year} {2003})\BibitemShut {NoStop}%
\bibitem [{\citenamefont {W\'ojcik}\ \emph {et~al.}(2018)\citenamefont
  {W\'ojcik}, \citenamefont {Bertoni},\ and\ \citenamefont
  {Goldoni}}]{Wojcik18}%
  \BibitemOpen
  \bibfield  {author} {\bibinfo {author} {\bibfnamefont {P.}~\bibnamefont
  {W\'ojcik}}, \bibinfo {author} {\bibfnamefont {A.}~\bibnamefont {Bertoni}},\
  and\ \bibinfo {author} {\bibfnamefont {G.}~\bibnamefont {Goldoni}},\
  }\bibfield  {title} {\bibinfo {title} {Tuning Rashba spin-orbit coupling in
  homogeneous semiconductor nanowires},\ }\href
  {https://doi.org/10.1103/PhysRevB.97.165401} {\bibfield  {journal} {\bibinfo
  {journal} {Phys. Rev. B}\ }\textbf {\bibinfo {volume} {97}},\ \bibinfo
  {pages} {165401} (\bibinfo {year} {2018})}\BibitemShut {NoStop}%
\bibitem [{\citenamefont {Wójcik}\ \emph {et~al.}(2019)\citenamefont
  {Wójcik}, \citenamefont {Bertoni},\ and\ \citenamefont
  {Goldoni}}]{Wojcik19}%
  \BibitemOpen
  \bibfield  {author} {\bibinfo {author} {\bibfnamefont {P.}~\bibnamefont
  {Wójcik}}, \bibinfo {author} {\bibfnamefont {A.}~\bibnamefont {Bertoni}},\
  and\ \bibinfo {author} {\bibfnamefont {G.}~\bibnamefont {Goldoni}},\
  }\bibfield  {title} {\bibinfo {title} {Enhanced Rashba spin-orbit coupling in
  core-shell nanowires by the interfacial effect},\ }\href
  {https://doi.org/10.1063/1.5082602} {\bibfield  {journal} {\bibinfo
  {journal} {Applied Physics Letters}\ }\textbf {\bibinfo {volume} {114}},\
  \bibinfo {pages} {073102} (\bibinfo {year} {2019})}
 \BibitemShut {NoStop}%
\bibitem [{\citenamefont {W\'ojcik}\ \emph {et~al.}(2021)\citenamefont
  {W\'ojcik}, \citenamefont {Bertoni},\ and\ \citenamefont
  {Goldoni}}]{Wojcik21}%
  \BibitemOpen
  \bibfield  {author} {\bibinfo {author} {\bibfnamefont {P.}~\bibnamefont
  {W\'ojcik}}, \bibinfo {author} {\bibfnamefont {A.}~\bibnamefont {Bertoni}},\
  and\ \bibinfo {author} {\bibfnamefont {G.}~\bibnamefont {Goldoni}},\
  }\bibfield  {title} {\bibinfo {title} {Anisotropy of the spin-orbit coupling
  driven by a magnetic field in InAs nanowires},\ } {\bibfield  {journal} {\bibinfo
   {journal} {Phys. Rev. B}\ }\textbf {\bibinfo {volume} {103}},\ \bibinfo
  {pages} {085434} (\bibinfo {year} {2021})}\BibitemShut {NoStop}%
\bibitem [{\citenamefont {Sitek}\ \emph {et~al.}(2018)\citenamefont {Sitek},
  \citenamefont {Urbaneja~Torres}, \citenamefont {Torfason}, \citenamefont
  {Gudmundsson}, \citenamefont {Bertoni},\ and\ \citenamefont
  {Manolescu}}]{Sitek18}%
  \BibitemOpen
  \bibfield  {author} {\bibinfo {author} {\bibfnamefont {A.}~\bibnamefont
  {Sitek}}, \bibinfo {author} {\bibfnamefont {M.}~\bibnamefont
  {Urbaneja~Torres}}, \bibinfo {author} {\bibfnamefont {K.}~\bibnamefont
  {Torfason}}, \bibinfo {author} {\bibfnamefont {V.}~\bibnamefont
  {Gudmundsson}}, \bibinfo {author} {\bibfnamefont {A.}~\bibnamefont
  {Bertoni}},\ and\ \bibinfo {author} {\bibfnamefont {A.}~\bibnamefont
  {Manolescu}},\ }\bibfield  {title} {\bibinfo {title} {Excitons in
  core-shell nanowires with polygonal cross sections},\ }\href@noop {}
  {\bibfield  {journal} {\bibinfo  {journal} {Nano Letters}\ }\textbf {\bibinfo
  {volume} {18}},\ \bibinfo {pages} {2581} (\bibinfo {year}
  {2018})}\BibitemShut {NoStop}%
\bibitem [{\citenamefont {Sitek}\ \emph {et~al.}(2016)\citenamefont {Sitek},
  \citenamefont {Thorgilsson}, \citenamefont {Gudmundsson},\ and\ \citenamefont
  {Manolescu}}]{Sitek16}%
  \BibitemOpen
  \bibfield  {author} {\bibinfo {author} {\bibfnamefont {A.}~\bibnamefont
  {Sitek}}, \bibinfo {author} {\bibfnamefont {G.}~\bibnamefont {Thorgilsson}},
  \bibinfo {author} {\bibfnamefont {V.}~\bibnamefont {Gudmundsson}},\ and\
  \bibinfo {author} {\bibfnamefont {A.}~\bibnamefont {Manolescu}},\ }\bibfield
  {title} {\bibinfo {title} {Multi-domain electromagnetic absorption of
  triangular quantum rings},\ }\href@noop {} {\bibfield  {journal} {\bibinfo
  {journal} {Nanotechnology}\ }\textbf {\bibinfo {volume} {27}},\ \bibinfo
  {pages} {225202} (\bibinfo {year} {2016})}\BibitemShut {NoStop}%
\bibitem [{\citenamefont {Sitek}\ \emph {et~al.}(2019)\citenamefont {Sitek},
  \citenamefont {Torres},\ and\ \citenamefont {Manolescu}}]{Sitek19}%
  \BibitemOpen
  \bibfield  {author} {\bibinfo {author} {\bibfnamefont {A.}~\bibnamefont
  {Sitek}}, \bibinfo {author} {\bibfnamefont {M.~U.}\ \bibnamefont {Torres}},\
  and\ \bibinfo {author} {\bibfnamefont {A.}~\bibnamefont {Manolescu}},\
  }\bibfield  {title} {\bibinfo {title} {Corner and side localization of
  electrons in irregular hexagonal semiconductor shells},\ }\href
  {https://doi.org/10.1088/1361-6528/ab37a1} {\bibfield  {journal} {\bibinfo
  {journal} {Nanotechnology}\ }\textbf {\bibinfo {volume} {30}},\ \bibinfo
  {pages} {454001} (\bibinfo {year} {2019})}\BibitemShut {NoStop}%
\bibitem [{\citenamefont {Sprung}\ \emph {et~al.}(1992)\citenamefont {Sprung},
  \citenamefont {Wu},\ and\ \citenamefont {Martorell}}]{Sprung}%
  \BibitemOpen
  \bibfield  {author} {\bibinfo {author} {\bibfnamefont {D.~W.~L.}\
  \bibnamefont {Sprung}}, \bibinfo {author} {\bibfnamefont {H.}~\bibnamefont
  {Wu}},\ and\ \bibinfo {author} {\bibfnamefont {J.}~\bibnamefont
  {Martorell}},\ }\bibfield  {title} {\bibinfo {title} {Understanding quantum
  wires with circular bends},\ }\href
  {https://doi.org/http://dx.doi.org/10.1063/1.350689} {\bibfield  {journal}
  {\bibinfo  {journal} {Journal of Applied Physics}\ }\textbf {\bibinfo
  {volume} {71}},\ \bibinfo {pages} {515} (\bibinfo {year} {1992})}\BibitemShut
  {NoStop}%
\bibitem [{\citenamefont {Ferrari}\ \emph {et~al.}(2009)\citenamefont
  {Ferrari}, \citenamefont {Goldoni}, \citenamefont {Bertoni}, \citenamefont
  {Cuoghi},\ and\ \citenamefont {Molinari}}]{Ferrari09b}%
  \BibitemOpen
  \bibfield  {author} {\bibinfo {author} {\bibfnamefont {G.}~\bibnamefont
  {Ferrari}}, \bibinfo {author} {\bibfnamefont {G.}~\bibnamefont {Goldoni}},
  \bibinfo {author} {\bibfnamefont {A.}~\bibnamefont {Bertoni}}, \bibinfo
  {author} {\bibfnamefont {G.}~\bibnamefont {Cuoghi}},\ and\ \bibinfo {author}
  {\bibfnamefont {E.}~\bibnamefont {Molinari}},\ }\bibfield  {title} {\bibinfo
  {title} {Magnetic states in prismatic core multishell nanowires},\
  }\href@noop {} {\bibfield  {journal} {\bibinfo  {journal} {Nano Letters}\
  }\textbf {\bibinfo {volume} {9}},\ \bibinfo {pages} {1631} (\bibinfo {year}
  {2009})}\BibitemShut {NoStop}%
\bibitem [{\citenamefont {Ballester}\ \emph {et~al.}(2012)\citenamefont
  {Ballester}, \citenamefont {Planelles},\ and\ \citenamefont
  {Bertoni}}]{Ballester12}%
  \BibitemOpen
  \bibfield  {author} {\bibinfo {author} {\bibfnamefont {A.}~\bibnamefont
  {Ballester}}, \bibinfo {author} {\bibfnamefont {J.}~\bibnamefont
  {Planelles}},\ and\ \bibinfo {author} {\bibfnamefont {A.}~\bibnamefont
  {Bertoni}},\ }\bibfield  {title} {\bibinfo {title} {Multi-particle states of
  semiconductor hexagonal rings: Artificial benzene},\ }\href@noop {}
  {\bibfield  {journal} {\bibinfo  {journal} {Journal of Applied Physics}\
  }\textbf {\bibinfo {volume} {112}},\ \bibinfo {eid} {104317} (\bibinfo {year}
  {2012})}\BibitemShut {NoStop}%
\bibitem [{\citenamefont {Klausen}\ \emph {et~al.}(2020)\citenamefont
  {Klausen}, \citenamefont {Sitek}, \citenamefont {Erlingsson},\ and\
  \citenamefont {Manolescu}}]{Klausen20}%
  \BibitemOpen
  \bibfield  {author} {\bibinfo {author} {\bibfnamefont {K.~O.}\ \bibnamefont
  {Klausen}}, \bibinfo {author} {\bibfnamefont {A.}~\bibnamefont {Sitek}},
  \bibinfo {author} {\bibfnamefont {S.~I.}\ \bibnamefont {Erlingsson}},\ and\
  \bibinfo {author} {\bibfnamefont {A.}~\bibnamefont {Manolescu}},\ }\bibfield
  {title} {\bibinfo {title} {Majorana zero modes in nanowires with combined
  triangular and hexagonal geometry},\ }\href
  {https://doi.org/10.1088/1361-6528/ab932e} {\bibfield  {journal} {\bibinfo
  {journal} {Nanotechnology}\ }\textbf {\bibinfo {volume} {31}},\ \bibinfo
  {pages} {354001} (\bibinfo {year} {2020})}\BibitemShut {NoStop}%
\bibitem [{\citenamefont {Bringer}\ and\ \citenamefont
  {Sch\"apers}(2011)}]{Bringer11}%
  \BibitemOpen
  \bibfield  {author} {\bibinfo {author} {\bibfnamefont {A.}~\bibnamefont
  {Bringer}}\ and\ \bibinfo {author} {\bibfnamefont {T.}~\bibnamefont
  {Sch\"apers}},\ }\bibfield  {title} {\bibinfo {title} {Spin precession and
  modulation in ballistic cylindrical nanowires due to the Rashba effect},\
  }\href {https://doi.org/10.1103/PhysRevB.83.115305} {\bibfield  {journal}
  {\bibinfo  {journal} {Phys. Rev. B}\ }\textbf {\bibinfo {volume} {83}},\
  \bibinfo {pages} {115305} (\bibinfo {year} {2011})}\BibitemShut {NoStop}%
\bibitem [{\citenamefont {Sitek}\ \emph {et~al.}(2017)\citenamefont {Sitek},
  \citenamefont {Urbaneja~Torres}, \citenamefont {Torfason}, \citenamefont
  {Gudmundsson},\ and\ \citenamefont {Manolescu}}]{Sitek17ICTON}%
  \BibitemOpen
  \bibfield  {author} {\bibinfo {author} {\bibfnamefont {A.}~\bibnamefont
  {Sitek}}, \bibinfo {author} {\bibfnamefont {M.}~\bibnamefont
  {Urbaneja~Torres}}, \bibinfo {author} {\bibfnamefont {K.}~\bibnamefont
  {Torfason}}, \bibinfo {author} {\bibfnamefont {V.}~\bibnamefont
  {Gudmundsson}},\ and\ \bibinfo {author} {\bibfnamefont {A.}~\bibnamefont
  {Manolescu}},\ }\bibfield  {title} {\bibinfo {title} {Controlled Coulomb
  effects in core-shell quantum rings},\ }\bibfield  {journal} {\bibinfo
  {journal} {in \textit{Proceedings of the 19th International Conference on Transparent
  Optical Networks (ICTON 2017)}}\ }
  (\bibinfo IEEE, Piscataway, NJ, {2017})\BibitemShut {NoStop}%
\bibitem [{\citenamefont {Campos}\ \emph {et~al.}(2018)\citenamefont {Campos},
	\citenamefont {Faria~Junior}, \citenamefont {Gmitra}, \citenamefont
	{Sipahi},\ and\ \citenamefont {Fabian}}]{Campos18}%
\BibitemOpen
\bibfield  {author} {\bibinfo {author} {\bibfnamefont {T.}~\bibnamefont
		{Campos}}, \bibinfo {author} {\bibfnamefont {P.~E.}\ \bibnamefont
		{Faria~Junior}}, \bibinfo {author} {\bibfnamefont {M.}~\bibnamefont
		{Gmitra}}, \bibinfo {author} {\bibfnamefont {G.~M.}\ \bibnamefont {Sipahi}},\
	and\ \bibinfo {author} {\bibfnamefont {J.}~\bibnamefont {Fabian}},\
}\bibfield  {title} {\bibinfo {title} {Spin-orbit coupling effects in
		zinc-blende InSb and wurtzite InAs nanowires: Realistic calculations with
		multiband $\mathbf{k}\ifmmode\cdot\else\textperiodcentered\fi{}\mathbf{p}$
		method},\ }\href {https://doi.org/10.1103/PhysRevB.97.245402} {\bibfield
	{journal} {\bibinfo  {journal} {Phys. Rev. B}\ }\textbf {\bibinfo {volume}
		{97}},\ \bibinfo {pages} {245402} (\bibinfo {year} {2018})}\BibitemShut
{NoStop}%
\bibitem [{\citenamefont {Basari\'c}\ \emph {et~al.}(2025)\citenamefont
  {Basari\'c}, \citenamefont {Brajovi\'c}, \citenamefont {Behner},
  \citenamefont {Moors}, \citenamefont {Schaarman}, \citenamefont {Manolescu},
  \citenamefont {Juluri}, \citenamefont {Sanchez}, \citenamefont {Bae},
  \citenamefont {L\"uth}, \citenamefont {Gr\"utzmacher}, \citenamefont
  {Pawlis},\ and\ \citenamefont {Sch\"apers}}]{Basaric25}%
  \BibitemOpen
  \bibfield  {author} {\bibinfo {author} {\bibfnamefont {F.}~\bibnamefont
  {Basari\'c}}, \bibinfo {author} {\bibfnamefont {V.}~\bibnamefont
  {Brajovi\'c}}, \bibinfo {author} {\bibfnamefont {G.}~\bibnamefont {Behner}},
  \bibinfo {author} {\bibfnamefont {K.}~\bibnamefont {Moors}}, \bibinfo
  {author} {\bibfnamefont {W.}~\bibnamefont {Schaarman}}, \bibinfo {author}
  {\bibfnamefont {A.}~\bibnamefont {Manolescu}}, \bibinfo {author}
  {\bibfnamefont {R.}~\bibnamefont {Juluri}}, \bibinfo {author} {\bibfnamefont
  {A.~M.}\ \bibnamefont {Sanchez}}, \bibinfo {author} {\bibfnamefont {J.~H.}\
  \bibnamefont {Bae}}, \bibinfo {author} {\bibfnamefont {H.}~\bibnamefont
  {L\"uth}}, \bibinfo {author} {\bibfnamefont {D.}~\bibnamefont
  {Gr\"utzmacher}}, \bibinfo {author} {\bibfnamefont {A.}~\bibnamefont
  {Pawlis}},\ and\ \bibinfo {author} {\bibfnamefont {T.}~\bibnamefont
  {Sch\"apers}},\ }
 {\bibinfo {title} {Aharonov-Bohm and Altshuler-Aronov-Spivak oscillations in the quasi-ballistic regime in phase-pure GaAs/InAs core/shell nanowires},\
}\bibinfo P Phys. Rev. B \textbf{112}, 075301 (2025)  \BibitemShut {NoStop}%
\bibitem [{\citenamefont {Sitek}\ \emph {et~al.}(2025)\citenamefont {Sitek},
  \citenamefont {Erlingsson},\ and\ \citenamefont {Manolescu}}]{Zenodo_SOI}%
  \BibitemOpen
  \bibfield  {author} {\bibinfo {author} {\bibfnamefont {A.}~\bibnamefont
  {Sitek}}, \bibinfo {author} {\bibfnamefont {S.~I.}\ \bibnamefont
  {Erlingsson}},\ and\ \bibinfo {author} {\bibfnamefont {A.}~\bibnamefont
  {Manolescu}},\ }\bibfield  {title} {\bibinfo {title} {Spin-orbit interaction
  in tubular prismatic nanowires [Data set], Zenodo, 2025},\ }\href
  {https://doi.org/10.5281/zenodo.15730966} {https://doi.org/10.5281/zenodo.15730966} \BibitemShut {NoStop}%
\end{thebibliography}

%

\end{document}